\documentclass[a4paper,11pt]{article}
\pdfoutput=1 % if your are submitting a pdflatex (i.e. if you have
\usepackage{jcappub} % for details on the use of the package, please
\usepackage[T1]{fontenc} % if needed
\RequirePackage{bm,amsfonts,lineno,amsmath,amssymb,float,eurosym,latexsym,epsf,mathtools,cuted,makecell,array}
\usepackage[font={footnotesize,it}]{caption}
\usepackage{epstopdf}
\usepackage{color}
\usepackage{footnote}
\setcellgapes{4pt}%parameter for the spacing

\title{\boldmath Charged anisotropic compact star in $f(R,\mathcal{T})$ gravity: A minimal geometric deformation gravitational decoupling approach}

\author[a,1]{S. K. Maurya\note{Corresponding author} }
\author[b]{Francisco Tello-Ortiz}

\affiliation[a]{Department of Mathematical and Physical Sciences,
College of Arts and Science, University of Nizwa, Nizwa, Sultanate of Oman}
\affiliation[b]{Departamento de F\'isica,
Facultad de ciencias b\'asicas, Universidad de Antofagasta, Casilla 170, Antofagasta, Chile}

\emailAdd{sunil@unizwa.edu.om}
\emailAdd{francisco.tello@ua.cl}

\abstract{This article is devoted to the study of high dense charged anisotropic compact structures in the framework of $f(R,\mathcal{T})$ gravity theory. The principal aims of this investigation, regard the extension of the isotropic Durgapal-Fuloria model within the context of charged isotropic $f(R,\mathcal{T})$ solutions.  The second main goal of this work is to apply the gravitational decoupling via a minimal geometric deformation (MGD) scheme in $f(R,\mathcal{T})$ gravity. Finally, the third one is to derive an anisotropic version of the charged isotropic model previously obtained by applying gravitational decoupling technology. This MGD approach splits the system of equations into two separate sets, one corresponding with the f(R,$\mathcal{T}$)  gravity system and another corresponding to the anisotropic sector governed by the extra source $\theta_{\mu\nu}$.To address this task we have considered some ingredients: I) the f(R,$\mathcal{T}$) model corresponds to a linear functional of the Ricci's scalar and the trace of the energy-momentum tensor $\mathcal{T}$, specifically $ f(R,\mathcal{T}) =R+2\chi\mathcal{T}$, being $\chi$ a running coupling constant. II) The matter distribution is taken to be an isotropic fluid one, III) the Lagrangian matter $\mathcal{L}_{m}$ corresponds to the negative isotropic pressure \i.e, $-p$ and IV) in order to solve the $\theta$-sector, a suitable form for the decoupler function $f(r)$ has been imposed, respecting all the physical and mathematical requirements.
  Finally, to check and contrast our model an exhaustive mathematical, physical and graphical analysis is performed in order to show all the properties that characterize the compact structure. It is worth mentioning that when  $\chi=0$ Einstein's gravity theory results are recovered.}   


\begin{document}
\maketitle
\flushbottom

\section{Introduction}
Despite the fruitful understanding that general relativity (GR) theory has given us about the evolution of the Universe and its hidden secrets, the presence of dark components \i.e. dark matter and dark energy, introduce some challenges to this beautiful theory. In fact, GR predicts that a Universe dominated by radiation or matter accelerates in a negative way due to gravitational attraction. However, at present several astronomical observations  reveal that the Universe is expanding in an accelerated way ~\cite{r1,r2,r3,r4,r5,r6,r7,r8,r9,r10,r11,r12,r13,r14,r15}. Therefore, it suggests that GR theory needs to be modified in order to explain this phenomenon which is caused by dark energy. Since the pioneering work by Starobinsky ~\cite{r16} on cosmic inflation in the framework of $f(R)$ gravity (here $R$ is the Ricci scalar), It has become an active research field. Essentially this theory modifies the usual Einstein-Hilbert (EH) action by the introduction of a function $f(R)$ which can contain high order derivative terms (Lorentz invariance terms) e.g. $R^{2}$, $R_{\mu\nu}R^{\mu\nu}$, etc. As can be seen, this theory modifies only the geometrical sector, allowing in principle address the presence of dark components as purely geometrical effects. Actually, Modified gravity theories (MGTs) have become an active research area in the cosmological scenario. Taking into account these issues, Qadir et al. ~\cite{r17} reinforced the requirement of the modified relativistic dynamics and indicated that this modification may help to settle down the problems related to dark matter and quantum gravity. The most MGTs are the mentioned $f(R)$ ~\cite{r18,r19,r20,r21}, $f(R,\mathcal{T})$ ~\cite{r22,r23,r24,r25}  ($\mathcal{T}$ is the trace of energy-momentum tensor), f($\mathcal{G}$) ~\cite{r26} ($\mathcal{G}$ is the Gauss-Bonnet term) and f(R,$\mathcal{T},R_{\mu\nu}T^{\mu\nu}$) gravity ~\cite{r27,r28,r29,r30}. In recent times, Nojiri et al. ~\cite{r31} presented many mathematical techniques to understand the problems of cosmos regarding the bouncing cosmos. They argued that gravity mediated by $f(R)$ and f($\mathcal{G}$) theories could be used to understand several hidden secrets of our Universe. Many cosmological applications were discussed in the context of f(R,$\mathcal{T}$)  theory ~\cite{r32,r33,r34,r35,r36,r37,r38}. Some of them are the non-static line element for collapsing of a spherical body having anisotropic fluid ~\cite{r39}, the static spherical wormhole solutions found in ~\cite{r40,r41}. Moreover, perturbation techniques were used in the study of spherical stars ~\cite{r42}. The effects on gravitational lensing due to f(R,$\mathcal{T}$) gravity were discussed in ~\cite{r43}. Baffou et al. ~\cite{r44} employed perturbation on de-Sitter space-time and power-law models in order to explore some cosmic viability bounds.
Currently, great interest has been the study of the existence of collapsed structures within the framework of these MGTs ~\cite{r45,r46,r47,r48}. As was pointed out before, $f(R,\mathcal{T})$  gravity theory has been considered as the most promising one in cosmological applications. Proposed by Harko et.al. ~\cite{r22} this theory regards different forms of the function $f(R,\mathcal{T})$ and can be seen as a generalization of the $f(R)$ gravity, where the introduction of the  trace of the energy-momentum tensor would come from the consideration of quantum effects. In recent years this theory has been employed in the study of collapsed structures such as neutron and quark stars. Moraes et.al. ~\cite{r49} obtained for the first time the Tolman-Oppenheimer-Volkoff (TOV) equation, which describes the equilibrium of the compact objects. The research in the arena of compact structures within the framework of $f(R, \mathcal{T})$ gravity, concerns isotropic, anisotropic and charged fluids ~\cite{r50,r51,r52}. It is well known that anisotropic fluids i.e. unequal radial and tangential pressure $(p_{r}\neq p_{t})$ represent a more realistic scenario form the astrophysical point of view. Abandon the isotropy condition gives rise an intriguing phenomenon inside the stellar interior e.g. when $p_{t}>p_{r}$ the system experiences a repulsive force that counteracts the gravitational gradient (attractive otherwise), which allows the construction of more compact and massive objects ~\cite{r53}, increasing value of the surface redshift ~\cite{r54,r55} which is an important quantity that relates the mass and the radius of the star, improvement of system stability, etc. On the other hand, the inclusion of electric charge also helps to enhance many features of the stars. Ivanov ~\cite{r56} argued that when an isotropic fluid sphere is not well behaved from the physical point of view, the presence of electric charge introduces some ingredients that improve the behavior of the system e.g. due to the repulsive Coulomb's force the system is under a repulsive force that counterbalances the gravitational one (the same like in the case of positive anisotropy factor $\Delta\equiv p_{t}-p_{r}>0$ ), then the equilibrium of the system is enhanced, another important feature is related to the energy density associated with the  electric field, which has a significant role in producing the gravitational mass of the object. The inclusion of anisotropies and charge in stellar interiors has a long history, Ruderman \cite{r57} theoretical works are considered as the cornerstone by subsequences investigation \cite{r58,r59,r60,r61,r62,r63,r64,r65,r66,r67,r68,r69,r70,r71} in this direction. As Mak and Harko have argued ~\cite{r72}, anisotropy can arise in different contexts such as the existence of a solid core or by the presence of type $3A$ superfluid ~\cite{r73}, pion condensation ~\cite{r74} or different kinds of phase transitions ~\cite{r75}. Respect to the inclusion of electric charge Bonnor's ~\cite{r76} pioneering work opened the door to this interesting research area.

At present, several papers available in the literature address the study of compact objects describing anisotropic matter distributions ~\cite{r77} (and references contained therein). Recently, a simpler but powerful mechanism developed by Ovalle and his collaborators to introduce anisotropic behavior in the stellar matter ~\cite{Ovalle1,Ovalle2}, has been proved to be a versatile method. Originally, this method was developed to deform Schwarzschild's space-time ~\cite{Ovalle3,Ovalle4} in the Randall-Sundrum brane-world scenario ~\cite{Randall1,Randall2}. This method is known as gravitational decoupling by a minimal geometric deformation (MGD) approach. Basically, this scheme works by taking a spherically symmetric isotropic solution to Einstein field equations respecting the most general line element in curvature coordinates given by  
\begin{equation}\label{eq1.1}
ds^{2} = e^{\nu(r) } \, dt^{2}-e^{\lambda(r)} dr^{2} +r^{2}(d\theta ^{2} -\sin ^{2} \theta \, d\phi ^{2}),
\end{equation}
whose material content is modified by the inclusion of a new material fields as follows 
\begin{equation}\label{eq1}
T_{\mu\nu}=\tilde{T}_{\nu\mu}+\beta\theta_{\mu\nu}
\end{equation}
where $\tilde{T}_{\mu\nu}$ represents a perfect fluid energy-momentum tensor (this source is referred as the seed matter distribution), $\beta$ is a dimensionless constant and the new source $\theta$ contains new fields like a scalar, vector or tensor fields. The complete set of equations to be solved with the presence of this extra source is to intricate. To tackle it, a deformation is performed on the metric potentials (MGD corresponds to deform only one of them, usually realized on the radial component) as follows
\begin{eqnarray}\label{eq2}
e^{-\lambda(r)}\mapsto \xi(r)+\beta f(r),
\end{eqnarray}
where $f(r)$ is the deformation function (for a more detailed discussion see sec. 2 and references ~\cite{Ovalle5,Ovalle6,Ovalle7,Ovalle8,Ovalle9,Ovalle10,Ovalle11}). This deformation allows splitting the system of  equations into two decoupled sets, one corresponding to the isotropic sector and the second one corresponding to the $\theta$ sector. The first system is satisfied by an isotropic solution of Einstein equations, while $\theta$-system is solved imposing some restrictions on the theta components or some adequate expression for the deformation function $f(r)$. Some recent applications of this method in different contexts such as: extending isotropic solutions of Einstein fields equations to anisotropic domains ~\cite{Ovalle12,Ovalle13,Ovalle14,Ovalle15,Ovalle16},  charged anisotropic fluid spheres ~\cite{Ovalle17,Ovalle18}, cloud of strings ~\cite{Ovalle19}, cylindrical symmetry ~\cite{Ovalle20}, black holes in $3+1$ and $2+1$ dimensions ~\cite{Ovalle21,Ovalle22,Ovalle23}, the inverse problem ~\cite{Ovalle24}, interior solution in $2+1$ dimensions ~\cite{Ovalle25} and in the framework of higher dimensional ~\cite{Ovalle26} and modified gravity theories f($\mathcal{G}$) ~\cite{Ovalle27} and f(R) ~\cite{Ovalle28}, can be found. Moreover, the method was extended including deformation over both metric functions ~\cite{Ovalle29,Ovalle30}.

Following this direction, in this article we study the existences of charged anisotropic fluid distributions in the arena of $f(R,\mathcal{T})$ gravity theory, where the anisotropic behaviour into the stellar interior rises due to the extra matter field $\theta_{\mu\nu}$, introduced by gravitational decoupling via MGD grasp. Since, the $f(R,\mathcal{T})$ functional perchance a very complex combination of Ricci's scalar $R$ and the trace of the energy-momentum tensor $\mathcal{T}$, in this opportunity for the sake of simplicity we have considered a linear dependence on $R$ and $\mathcal{T}$. In particular $f(R,\mathcal{T})=R+2\chi\mathcal{T}$, where $\chi$ is a dimensionless running constant. This assumption on the $f(R,\mathcal{T})$ functional allows to split the intricate system of equations as required by MGD scheme \i.e, separate the global set of equations into the original $f(R,\mathcal{T})$ system and the corresponding one to the $\theta$-sector. Moreover, we want to stress that the inclusion of more complicated terms into the $f(R,\mathcal{T})$ functional introduce mathematical complications that make the problem intractable. However, if one wishes to consider a more complex model, in order to achieve gravitational decoupling one can rewrite a kind of effective Einstein tensor that contains the new terms. This way of proceeding has been taken into account in \cite{Ovalle27,Ovalle28}. Despite its simplicity the chosen model can be consolidated as Einstein gravity plus a running effective  cosmological constant, with the particularity that this running effective cosmological constant breaks down the minimal coupling matter principle. This is an important thing to keep in mind, since the breaking of the minimal coupling  matter principle between the gravitational interaction and the matter sector results in the non-conservation of the energy-momentum tensor. Consequently, the hydrostatic balance of the system is affected by remembering that the free divergence of the energy-momentum tensor leads to the Tolman-Oppenheimer-Volkoff (TOV) equilibrium equation. In addition, as a requirement of the gravitational decoupling method through MGD requires that once the systems are separated, Bianchi's identities should be satisfied for the respective sources \i.e, $\tilde{T}_{\mu\nu}$ (the seed matter distribution) and $\theta_{\mu\nu}$. This means that both sources only interact gravitationally. As stated before, $f(R,\mathcal{T})$ theory violates Bianchi's identities, however all the new contributions from the material sector $\mathcal{T}$ (in this case the geometric part remains the same like in GR) can be identified as part of an effective energy-momentum tensor which satisfies Bianchi's identities \i.e, $\nabla^{\mu}T^{(eff)}_{\mu\nu}$.

As we want to appreciate the effects of the $f(R,\mathcal{T})$ model into the $\theta$-sector, we have built our seed solution (which describes a charged isotropic fluid sphere) by using the well-known Durgapal-Fuloria \cite{Durgapal} radial metric potential $\xi(r)$ as ansatz. The particularity of our input is that the obtained electric field $E^{2}(r)$ and the temporal metric component $e^{\nu(r)}$ contain the dimensionless running constant $\chi$. This characteristic ensures that the modifications coming from $f(R,\mathcal{T})$ gravity theory go to $\theta$-sector after decoupling the sources. This is because after gravitational decoupling, the resulting systems of equations are connected by the temporal metric potential. The present procedure is plausible given that both theories, that is, $f(R,\mathcal{T})$ and GR theories share the same isotropy condition, which means that any solution to Einstein's field equations is also a solution within the framework of $f(R,\mathcal{T})$ gravity theory. Clearly, this is manifest only from the geometric point of view, since the material content of the fluid sphere is totally different. Another points to highlight in this study are the following ones: I) we have taken the Lagrangian $\mathcal{L}_{m}$ matter to be the negative of the isotropic pressure $-p$. This choice is stipulated by the statement that the action $S$ must be the integral of a scalar density. II) To close the system of equations related with the $\theta$ source we have imposed an adequate form of the decoupler function $f(r)$ respecting all the minimal requirements in order to describe a well behaved interior solution. III) The obtained charged anisotropic fluid sphere is joined in a smoothly way with Reissner-Nordstr\"{o}m exterior space-time to get the constant and physical parameters that characterize our model. It was possible due to the extra matter piece $\mathcal{T}$ is confined in the range $0\leq r\leq R$ \i.e, within the stellar distribution. Consequently the well known Israel-Darmois \cite{r78,r79} matching conditions can be employed to compute all the physical and constant parameters of the model. 
So, the article is organized as follows, in Sec. \ref{sec2} we develop the gravitational decoupling method through a minimal geometric deformation approach within the framework of $f(R,\mathcal{T})$ gravity, next in Sec. \ref{sec3} the main field equation for charged anisotropic matter distribution is presented. In Sec. \ref{sec4} the full model is obtained and its thermodynamic characterization is provided, Sec. \ref{sec5} presents the necessary and sufficient set of equations in order to determine all the constant parameters that characterize the solution. To do so, the junction condition between the interior geometry and the exterior space-time described by Reissner-Nordstr\"{o}m metric is performed. In secs. \ref{sec6} and \ref{sec7} are analyzed the main physical quantities such as effective density, effective radial and tangential pressure, anisotropy factor, electric field, and the equilibrium and stability conditions, respectively. Furthermore, the effect of the different parameters on the principal system quantities is discussed. Sec. \ref{sec8} discusses the behaviour of mass-radius ratio relation and its implications on the surface redshift, Sec. \ref{sec9} talks over the energy conditions and the necessary constraints on the energy-momentum tensor that describe the matter contained within the star. Finally, in Sec. \ref{sec10} some conclusions of the reported study are given.

%%%%%%%%%%%%%%%%%%%%%%%%%%%%%%%%%%%%%%%%%%%%%%%%%%%%

\section{Gravitational decoupling $f(R, \mathcal{T}$)   gravity formulation}\label{sec2}
The starting point, is the integral action given by
\begin{eqnarray}\label{1.1}
S=\frac{1}{16\pi}\int f(R,\mathcal{T})\sqrt{-g} d^{4}x+\int \mathcal{L}_{m}\sqrt{-g}d^{4}x +\beta\int \mathcal{L}_{X}\sqrt{-g} d^{4}x+\int \mathcal{L}_{e}\sqrt{-g} d^{4}x,
\end{eqnarray}
where relativistic geometrized units were employed i.e $c=G=1$. Here, $f(R,\mathcal{T})$ is an arbitrary function of the Ricci scalar $R$
and the trace $\mathcal{T}$ of the energy-momentum tensor $T_{\mu\,\nu}$, standing $\mathcal{L}_{m}$ the Lagrangian density matter fields, $\mathcal{L}_{X}$ the Lagrangian density of a new sector, not necessarily described by general relativity and $\mathcal{L}_{e}$ the Lagrangian electromagnetic field. This new sector can be seen as corrections to general relativity and can be interpreted as part of an effective energy-momentum tensor.   \\
Variation respect to the metric tensor $g_{\mu\nu}$ yields the following field equation
\begin{eqnarray}\label{1.2}
\left( R_{\mu\nu}- \nabla_{\mu} \nabla_{\nu} \right)f_R (R,\mathcal{T}) +\Box f_R (R,\mathcal{T})g_{\mu\nu} - \frac{1}{2} f(R,\mathcal{T})g_{\mu\nu} = 8\pi\left( T_{\mu\nu} +\beta\theta_{\mu\nu}\right) \nonumber \\ - f_\mathcal{T}(R,\mathcal{T})\, \left(T_{\mu\nu}  +\Theta_{\mu\nu}\right)+8\,\pi\,E_{\mu\,\nu}~~~~~
\end{eqnarray}

where $f_R (R,\mathcal{T})$ denote the partial derivative of $f (R,\mathcal{T})$ with respect to $R$ while $f_\mathcal{T} (R,\mathcal{T})$ is the partial derivative of $f (R,\mathcal{T})$ with respect to $\mathcal{T}$. $R_{\mu\nu}$ is the Ricci tensor, \\
${\Box \equiv \partial_{\mu}(\sqrt{-g} g^{\mu\nu} \partial_{\nu})/\sqrt{-g}}$ is the D'Alambert operator,
 and $\nabla_\mu$ represents the
covariant derivative which is associated with the Levi-Civita connection of metric tensor $g_{\mu\nu}$.

The stress energy tensor $T_{\mu\nu}$, electromagnetic tensor $E_{\mu\,\nu}$, the extra source $\theta_{\mu\nu}$ and $\Theta_{\mu\nu}$ are defined as follow,
\begin{eqnarray}
T_{\mu\nu}&=&g_{\mu\nu}\mathcal{L}_m-2\partial\mathcal{L}_m/\partial g^{\mu\nu},\\
E_{\mu\nu}&=&-g_{\mu\nu}\mathcal{L}_e+2\partial\mathcal{L}_e/\partial g^{\mu\nu},\\
\theta_{\mu\nu}&=&g_{\mu\nu}\mathcal{L}_{X}-2\partial\mathcal{L}_{X}/\partial g^{\mu\nu},\\
\Theta_{\mu\nu}&=& g^{\beta\beta}\delta T_{\beta\beta}\,/\,\delta g^{\mu\nu}.
\end{eqnarray}
The equation (\ref{1.2}) can be rearranged as follows
\begin{eqnarray}\label{eins}
G_{\mu\nu}=\frac{1}{f_{R}\left(R,\mathcal{T}\right)}\bigg[8\pi\left(T_{\mu\nu}+\beta\theta_{\mu\nu}\right)+\frac{1}{2}\left(f\left(R,\mathcal{T}\right)-Rf_{R}\left(R,\mathcal{T}\right)\right)g_{\mu\nu}   -\left(T_{\mu\nu}+\Theta_{\mu\nu}\right)f_{\mathcal{T}}\left(R,\mathcal{T}\right)\nonumber \\ -\left(g_{\mu\nu}\Box-\nabla_{\nu}\nabla_{\mu}\right)f_{R}\left(R,\mathcal{T}\right)+8\,\pi\,E_{\mu\,\nu}\bigg]. ~~~
\end{eqnarray}

where $G_{\mu\nu}$ corresponds to Einstein tensor. The corresponding conservation equation reads
\begin{eqnarray}\label{1.3}
\resizebox{1\hsize}{!}{$\nabla^{\mu}T_{\mu\nu}=\frac{f_\mathcal{T}(R,\mathcal{T})}{8\pi -f_\mathcal{T}(R,\mathcal{T})}\bigg[(T_{\mu\nu}+\Theta_{\mu\nu})\nabla^{\mu}\ln f_\mathcal{T}(R,\mathcal{T})  +\nabla^{\mu}\Theta_{\mu\nu}-\frac{1}{2}g_{\mu\nu}\nabla^{\mu}\mathcal{T} -\frac{8\pi}{f_{\mathcal{T}}\left(R,\mathcal{T}\right)} \big(\beta\,\nabla^{\mu}\theta_{\mu\nu}+\nabla^{\mu}E_{\mu\,\nu}\big)\bigg]$}.~~~~~
\end{eqnarray}

The equation (\ref{1.3}) shows that the stress-energy momentum tensor $T_{\mu\nu}$ in $f(R,T)$ gravity is not conserved, as in other theories of gravity. On the other hand, in the present study we are choosing the energy-momentum tensor $T_{\mu\nu}$ for the isotropic matter distribution of the form,
\begin{equation}\label{1.4}
T_{\mu\nu}=(\rho+p)u_\mu u_\nu-pg_{\mu\nu}
\end{equation}
 and corresponding electromagnetic field tensor is given as
 \begin{equation}\label{1.44}
E_{\mu\nu}=\frac{1}{4 \pi}(-F^{m}_{\mu}F_{\nu\,m} + \frac{1}{4}{g_{\mu\nu}} F_{\gamma\,n}F^{\gamma\,n})
\end{equation}

where ${u_{\nu}}$ is the four velocity, satisfying $u_{\mu}u^{\mu}=-1$ and $u_{\nu}\nabla^{\mu}u_{\mu}=0$. Here, $\rho$ is matter density and $p$ the isotropic pressure, respectively.  

The anti-symmetric electromagnetic field tensor $F_{\mu \nu}$ in Eq. (\ref{1.44}) is defined as
\begin{eqnarray}
F_{\mu \nu}=\nabla_{\mu}\,A_{\nu}-\nabla_{\nu}\,A_{\mu}
\end{eqnarray}
which satisfies the Maxwell equations,
\begin{eqnarray}
F_{\mu \nu,\gamma} + F_{\nu \gamma,\mu}+F_{\gamma \mu, \nu}=0
\end{eqnarray}
and
\begin{equation}
 F^{ik};k = 4\pi\,J^i  \label{8a}
\end{equation}
where, the electromagnetic four current vector $j^i$ is defined as
\begin{eqnarray}
J^{i}=\frac{\sigma}{\sqrt{g_{44}}}\,\frac{dx^i}{dx^4}=\sigma\,v^i, \label{8b}
\end{eqnarray}
with charge density $\sigma=e^{\nu/2}\,J^{0}(r)$. For static matter distribution the only non-zero component
of the four-current is $J^4$. Because of spherical symmetry, the four-current component is only a function of radial distance, r . The only non-vanishing components of electromagnetic
field tensor are $F^{41}$ and $F^{14}$, related by $F^{41}=-F^{14}$, which describe the radial component of the electric field. From Eqs.
(\ref{8a}) and (\ref{8b}), one obtains the following expression for the component of the electric field,
\begin{eqnarray}
F^{41} = - F^{14}=\frac{q}{r^2}\,e^{-(\nu+\lambda)/2}
\end{eqnarray}
The $q(r)$ describes the effective charge
contained within the sphere of radius r , then it can be defined by the relativistic Gauss law  and corresponding electric field $E$ as,
\begin{eqnarray}
q(r)&=&4\,\pi\int^{r}_{0}{\sigma\,r^2\,e^{\lambda/2} dr}=r^2\,\sqrt{-F_{14}\,F^{14}}\\
E^2&=&-F_{14}\,F^{14}=\frac{q^2}{r^4}
\end{eqnarray}

In our study we will consider throughout the Lagrangian matter  $\mathcal{L}_m=-p$. Then for  $\Theta_{\mu\nu}$, we obtain $\Theta_{\mu\nu}=-2T_{\mu\nu}-p g_{\mu\nu}$.

In order to determine the effective stress-energy momentum tensor for modified theory of gravity  we choose the linear functional form of $f(R,\mathcal{T})$ as follows
\begin{eqnarray}
f(R,\mathcal{T})=R+2\chi\mathcal{T},
\label{frt}
\end{eqnarray}
where $\chi$ is a coupling constant. The above linear functional has been used successfully in other different $f(R,\mathcal{T})$ gravity models. By plugging the value $f(R,\mathcal{T})$ from Eq. (\ref{frt}) in Eq. (\ref{eins}) we obtain
\begin{eqnarray}\label{1.5}
G_{\mu\nu}=8\pi \left(T_{\mu\nu}+\beta\theta_{\mu\nu}\right)+\chi \mathcal{T}g_{\mu\nu}+2\chi(T_{\mu\nu}+p g_{\mu\nu}) + 8\pi\,E_{\mu\nu}=8\pi T^{eff}_{\mu\nu}+8\pi\,E_{\mu\nu}.~~~
\end{eqnarray}

Here $G_{\mu\nu}$ and $E_{\mu\nu}$ represent the Einstein tensor and electromagnetic tensor respectively while $T_{\mu\nu}$ denotes the energy momentum tensor corresponding to perfect fluid distribution. Here effective energy-momentum tensor $T^{eff}_{\mu\nu}$  is combination of energy-momentum tensor due  to the matter geometry coupling $\chi$ and extra source tensor $\theta_{\mu\nu}$ (which generates the anisotropy in $T^{eff}_{\mu\nu}$) which can be written as
\begin{eqnarray}\label{1.5a}
T^{(eff)}_{\mu\nu}= \tilde{T}_{\mu\nu} +\beta\,\theta_{\mu\nu}~~
\end{eqnarray}
 where,
 \begin{eqnarray}
\tilde{T}_{\mu \nu}= {T}_{\mu\nu}\,\left(1+\frac{\chi}{4\pi}\right)+\frac{\chi}{8\pi}(\mathcal{T}+2p)g_{\mu\nu}.\label{1.5aa}
 \end{eqnarray}

 It is note that ${T}_{\mu \nu}$ is energy momentum tensor corresponds to stress energy tensor  which is given  by Eq. (11).
By inserting the value of $f(R,\mathcal{T})=R+2\chi\mathcal{T}$ in Eq.(\ref{1.3}) we get
\begin{eqnarray}\label{1.6a}
\nabla^{\mu}T_{\mu\nu}=-\frac{1}{2\,\left(4\pi+\chi\right)}\chi\bigg[g_{\mu\nu}\nabla^{\mu}\mathcal{T}+2\,\nabla^{\mu}(p g_{\mu\nu})+\beta\frac{8\pi}{\chi}\nabla^{\mu}\theta_{\mu\nu}+\frac{8\pi}{\chi}\nabla^{\mu}E_{\mu\nu}\bigg].
\end{eqnarray}
By using the Eqs.(\ref{1.5a}) and (\ref{1.6a}) we can write,
\begin{eqnarray}
\nabla^{\mu}T^{(eff)}_{\mu\nu}=0\label{1.6aa}
\end{eqnarray}
On the other hand we define the effective tensor $T^{(eff)}_{\mu\nu}$ for anisotropic matter distribution as
\begin{eqnarray}
T^{(eff)}_{\mu\nu}=(\rho^{eff} + p^{eff}_t)u_\mu u_\nu  + (p^{eff}_r - p^{eff}_t) v_\mu v_\nu- p^{eff}_t{g_{\mu\nu}}. ~~~\label{1.6bb}
\end{eqnarray}

Where ${\rho^{eff}}$, ${p}^{eff}_{r}$ and ${p}^{eff}_{t}$ represent the effective density, the effective radial pressure and the effective tangential pressure respectively. whereas $u_{\mu}$ and $v_{\nu}$ denote four-velocity and radial four-vector, respectively.

\section{Main equations for charged anisotropic matter distributions}\label{sec3}
Let us consider the spacetime being static and spherically symmetric, which describes the interior of the object can be written in the following form
\begin{equation}\label{eq1a}
ds^{2} = e^{\nu(r) } \, dt^{2}-e^{\lambda(r)} dr^{2} +r^{2}(d\theta ^{2} -\sin ^{2} \theta \, d\phi ^{2}),
\end{equation}

Since the effective energy tensor $T^{(eff)}_{\mu\nu}$ involves the extra source $\theta_{\mu\nu}$ which will clearly generates the anisotropic pressure within the effective matter distribution. Hence, using Using Eqs. (\ref{1.5}) and (\ref{1.6bb}) together with line element (\ref{eq1a}) we determine the Einstein-Maxwell field equations for the spherically symmetric charged anisotropic stellar system as

\begin{eqnarray}\label{effectivedensity}
8\pi\,\rho^{eff} +\frac{q^2}{r^4} &=&\frac{1}{r^2}-e^{-\lambda}\left(\frac{1}{r^2}-\frac{\lambda^{\prime}}{r}\right)\\\label{effectiveradialpressure}~~~
8\pi\, {p}^{eff}_{r}-\frac{q^2}{r^4}&=&-\frac{1}{r^2}+e^{-\lambda}\left(\frac{1}{r^2}+\frac{\nu^{\prime}}{r}\right)\\\label{effectivetangentialpressure}
8\pi\,{p}^{eff}_{t}+\frac{q^2}{r^4}&=&\frac{1}{4}e^{-\lambda}\left(2\nu^{\prime\prime}+\nu^{\prime2}-\lambda^{\prime}\nu^{\prime}+2\frac{\nu^{\prime}-\lambda^{\prime}}{r}\right).~~~~~
\end{eqnarray}
The primes denote differentiation with respect to the radial coordinate $r$. Using the Eqs.(\ref{1.5a}) and (\ref{1.5aa}) the effective quantities like effective density (${\rho^{eff}}$), effective radial pressure (${p}^{eff}_{r}$) and effective tangential pressure (${p}^{eff}_{t}$) can be written in terms of isotropic pressure ($p$), density ($\rho$) for isotropic matter distribution and $\theta$ components as,
\begin{eqnarray}\label{effecrho}
{\rho^{eff}}&=&\tilde{\rho}+\beta\, \theta^{ 0}_{0}\\\label{effecpr}
{p}^{eff}_{r}&=&\tilde{p}-\beta\, \theta^{1}_{1}\\ \label{effecpt}
{p}^{eff}_{t}&=& \tilde{p}-\beta\, \theta^{2}_{2},
\end{eqnarray}
where,
\begin{eqnarray}
\tilde{\rho}&=&\rho+\frac{\chi}{8\,\pi}(3\rho-p),\\
\tilde{p}&=&p -\frac{\chi}{8\,\pi}(\rho-3 p)
\end{eqnarray}

The presence of the $\theta$-term in above system clearly introduce an anisotropy if $\theta^{1}_{1}\neq \theta^{2}_{2}$. Thus the effective anisotropy is defined as
\begin{equation}\label{anisotropy}
\Delta^{eff}\equiv {p}^{eff}_{t}-{p}^{eff}_{r}=\beta\left(\theta^{1}_{1}- \theta^{2}_{2}\right).
\end{equation}
\section{Solution of Einstein-Maxwell field Equations via Gravitational decoupling approach:}\label{sec4}

Solve the system of equations (\ref{effectivedensity})-(\ref{effectivetangentialpressure}) is not an easy task. In order to tackle it we will employ the gravitational decoupling via the MGD approach. This method consists in deforming the metric potentials $e^{\nu(r)}$ and  $e^{\lambda(r)}$ through a linear mapping given by
\begin{eqnarray}\label{deformationnu}
e^{\nu(r)}&\mapsto& e^{\nu(r)}+\beta h(r) \\ \label{deformationlambda}
e^{-\lambda(r)}&\mapsto& \xi(r)+\beta f(r),
\end{eqnarray}
where $h(r)$ and $f(r)$ are the corresponding deformations. It's worth mentioning that the foregoing deformations are purely radial functions, this feature ensures the spherical symmetry of the solution. The so called MGD corresponds to set $h(r)=0$ or $f(r)=0$, in this case the deformation will be done only on the radial component, remaining the temporal one unchanged (it corresponds to set $h(r)=0$). Then the anisotropic sector $\theta_{\mu\nu}$ is effectively contained in the radial deformation (\ref{deformationlambda}). By inserting Eq. (\ref{deformationlambda}) into the system of equations  (\ref{effectivedensity})-(\ref{effectivetangentialpressure}) and using decoupled systems of equations (\ref{effecrho})-(\ref{effecpt}). The first set of equations corresponds to $\beta=0$, it means charged perfect fluid matter distribution, as
\begin{eqnarray}
 {8\,\pi}\,\tilde{\rho}+\frac{q^2}{r^4} &=&  \bigg[\frac{1}{r^{2}}-\frac{\xi}{r^{2}}-\frac{\xi^{\prime}}{r}\bigg] \label{dr1} \\
{8\,\pi}\,\tilde{p}-\frac{q^2}{r^4}&=& \bigg[\xi\left(\frac{1}{r^{2}}+\frac{\nu^{\prime}}{r}\right)-\frac{1}{r^{2}}\bigg]\label{p2}\\
{8\,\pi}\,\tilde{p}+\frac{q^2}{r^4}&=& \bigg[\frac{\xi}{4}\left(2\nu^{\prime\prime}+\nu^{\prime2}+2\frac{\nu^{\prime}}{r}\right)+\frac{\xi^{\prime}}{4}\left(\nu^{\prime}+\frac{2}{r}\right)\bigg].~~~~\label{p3}
\end{eqnarray}
along with the conservation equation
\begin{eqnarray}
\nabla^{\mu}T_{\mu\nu}=\frac{-\chi}{2\,\left(4\pi+\chi\right)}\bigg[g_{\mu\nu}\nabla^{\mu}\mathcal{T}+2\,\nabla^{\mu}(p g_{\mu\nu})+\frac{8\pi}{\chi}\nabla^{\mu}E_{\mu\nu}\bigg].~~~~~~
\end{eqnarray}

 The other set of equations corresponds to the $\theta$ sector as,
\begin{eqnarray}\label{one}
8\pi\theta^{0}_{0}&=&-\frac{f}{r^{2}}-\frac{f^{\prime}}{r}
\\
8\pi\theta^{1}_{1}&=&-f\left(\frac{1}{r^{2}}+\frac{\nu^{\prime}}{r}\right)  \\  \label{two}
8\pi\theta^{2}_{2}&=&-\frac{f}{4}\left(2\nu^{\prime\prime}+\nu^{\prime2}+2\frac{\nu^{\prime}}{r}\right)-\frac{f^{\prime}}{4}\left(\nu^{\prime}+\frac{2}{r}\right). \label{three}
\end{eqnarray}
The corresponding conservation equation $\nabla^{\nu}\theta_{\mu\nu}=0$ yields to
\begin{eqnarray}
\left(\theta^{1}_{1}\right)^{\prime}-\frac{\nu^{\prime}}{2}\left(\theta^{0}_{0}-\theta^{1}_{1}\right)-\frac{2}{r}\left(\theta^{2}_{2}-\theta^{1}_{1}\right)=0,\label{conservationtheta}
\end{eqnarray}
This expression is a linear combination of the quasi-Einstein equations. At this point it is remarkable to note that both the isotropic charged and the anisotropic sectors $\theta_{\mu\nu}$ are individually conserved, it means that both systems interact only gravitationally.

\subsection{{Solution of Einstein-Maxwell field Equations (\ref{dr1})- (\ref{p3}) in $f(R, \mathcal{T})$ gravity:}}

In order to solve the Equation (\ref{dr1}), (\ref{p2}) and (\ref{p3}), we use the isotropy condition in Eqs. (\ref{p2}) and (\ref{p3}) which gives the second order differential equation as,
\begin{eqnarray}
{\xi}\left(\frac{\nu^{\prime\prime}}{2}+\frac{\nu^{\prime2}}{4}-\frac{\nu^{\prime}}{2\,r}-\frac{1}{r^2}\right)+\frac{\xi^{\prime}\nu^{\prime}}{4}+\frac{2\,\xi^{\prime}}{r}-\frac{1}{r^2}=\frac{2q^2}{r^4}~~~~~~\label{diff}
\end{eqnarray}
The above equation having three unknown $\nu$, $\xi$ and $\frac{q^2}{r^4}$. In order to obtain a general solution of above Eq.(\ref{diff}) in $f(R,\mathcal{T})$ gravity, we take the anstaz of gravitational potential $\xi$ due to Durgapal-flouria \cite{Durgapal} as
\begin{eqnarray}
\xi&=&\frac{7-10A\,r^2-A^2\,r^4}{7 + 14 Ar^2 + 7A^2\,r^4} \label{garv1}
\end{eqnarray}
where $A$ is a positive constant. We note that the ansatz for the gravitational potential $g_{rr}$, given by (\ref{garv1}), was discovered by Durgapal-flouria \cite{Durgapal} to construct a physically viable relativistic perfect fluid solution for superdense star model. The selection of this gravitational potential $g_{rr}$ is physically well motivated (especially as the energy density must be non-singular positive and decreasing outward). In the past, this potential (\ref{garv1}) has been used to develop a viable stellar model for charged and anisotropic matter distribution in General Relativity. Moreover, the above the gravitational potential (\ref{garv1}) is also free from a singularity, positive and finite at center $r=0$ of a stellar model and increasing monotonically towards the boundary of the stellar model. It is note that if electric charge $q=0$ then $\nu=4\,\ln(1+Ar^2)$ will be a particular solution of Eq.(\ref{diff}) for isotropic matter distribution. By keeping this point in our mind, we construct the expression for the electric charge function $q$ of the form

\begin{eqnarray}
\frac{q^2}{r^4}&=&\frac{4A^2r^2\,\big[(1 +\chi)^2\,(5 + Ar^2) + \alpha^2\,(7 -15Ar^2 - 2A^2r^4) + 2\alpha(1 + \chi)\,(-6 + 7Ar^2 + A^2r^4)\big]}{7(1 + Ar^2)^3\,(1 + \chi + \alpha\,Ar^2)^2}.~~~~~~~~\label{charge}
\end{eqnarray}

The above expression of electric charge involves two free parameters $\alpha$ and $\chi$ which will generate a class of charged isotropic solution for $f(R,\mathcal{T})$ gravity system.  Form above Eq.(\ref{charge}), It is obvious that the electric charge is zero at centre $r=0$ which shows that $q$ is free from singularity at $r=0$. It is interesting that the electric charge vanishes throughout within the compact stellar object for $\chi=0$ and $\alpha=1$ and gives a Durgapal-Fluria \cite{Durgapal} perfect fluid solution for the compact object in General relativity. On the other hand, if $\alpha=1$, then the presence of electric charge inside the matter distribution is only due to coupling parameter $\chi$ present in the system. The other details of the electric charge function $q$ are given in Sec. VI. \\

After plugging the gravitational potential $\xi$ and and electric charge function $\frac{q^2}{r^4}$ from Eqs.(\ref{garv1}) and (\ref{charge}) with the transformation $\nu=2\ln Y$ into Eq.(\ref{diff}) we get,

\begin{equation}
 \resizebox{1\hsize}{!}{$Y^{\prime\prime}-\frac{ (-7 - 21 A r^2 + 19 A^2 r^4 + A^3 r^6)}{r\,(1 + A r^2)  (-7 + 10 A r^2 + A^2 r^4)}\, Y^{\prime}-\frac{8\, \alpha\,A^2\,r^3\,\big[-4 (1 + \chi)\,(-3 + c r^2) +\alpha\, (-7 + 15 A r^2 + 7 A^2 r^4 + A^3 r^6)] }{r (1 + A r^2) (1 + \chi + \alpha  r^2)^2\,(-7 + 10 A r^2 + A^2 r^4)}\,Y=0$}.
 \label{Diff2}
\end{equation}

It is note that the value $Y=(1+\chi+\alpha\,A r^2)^2$ satisfy the above differential Eq.(\ref{Diff2}) which implies that this value of $Y$ represents a particular solution of Eq.(\ref{Diff2}). Then most general solution of Eq.(\ref{Diff2}) is given (using the change of dependent variable method) as,

\begin{equation}
\resizebox{0.97\hsize}{!}{$Y(r)=Y_{p}(r)\Bigg[C+D \Bigg(\frac{{Y_1(r)}\,\big[{Y_2}(r)+Y_3(r)\big]}{3\,(g_{\chi\,\alpha})^3\, (1 + \chi + \alpha Ar^2)^3} +\frac{4\,f_{\alpha\chi}}{\big(g_{\chi\alpha}\big)^{7/2}}\,\Large{\ln}\bigg[\frac{\alpha\,(g_{\chi\alpha})^3\,\left[Y_{4}(r) + \chi Ar^2\sqrt{g_{\chi\alpha}}\,\,{Y_1(r)}\right]}{2\,\sqrt{g_{\chi\alpha}}\,f_{\alpha\chi}\,\,(1 + \chi + \alpha Ar^2)}\bigg]\Bigg)\Bigg]$}. 
\label{Sol1}
\end{equation}
where $C$ and $D$ are arbitrary constant of integration and expression of the used coefficients are as follows:  \\
$Y_{p}(r)=(1+\chi+\alpha\,Ar^2)^2$,~~$Y_1(r)=\sqrt{7- 10 Ar^2 - A^2r^4}$,~~~$g_{\chi\alpha}=7 \alpha^2 + 10 \alpha (1 +\chi) - (1 + \chi)^2$,\\\\~~ ${Y_2}(r)=3 (1 + \chi)^5 + 3 \alpha (1 + \chi)^4 (-8 + Ar^2) + \alpha^2 (1 + \chi)^3  (185 + 4 Ar^2 + A^2r^4)$, \\
$Y_3(r)=a^3 (1 + \chi)^2 (-535 + 142 Ar^2 + 7 A^2r^4) + 3 a^4 (1 + \chi) (-84 - 316 Ar^2 + 21 A^2r^4)-\alpha^5 (49 + 161\,Ar^2 + 359\, A^2r^4)$,\\\\
$Y_4(r)= [5 + 5\,\chi + a (7 - 5 Ar^2) +Ar^2] $,~~$f_{\alpha\chi}=\big[163 \alpha^3 - 37 \alpha^2 (1 + \chi) + a (1 + \chi)^2 + (1 + \chi)^3\big]$.\\

By plugging the value of $\xi$, $q$ and $\nu=2\ln Y$ from Eqs. (\ref{garv1}), (\ref{charge}) and (\ref{Sol1}) into Eqs.(\ref{dr1}) and (\ref{p2}) we find the pressure ($\tilde{p}$) and $\tilde{\rho}$ in $f(R,\mathcal{T})$ gravity as

\begin{eqnarray}
&&\tilde{p}=\frac{4A\,Y_1(r)\big[D\,(1 + Ar^2)+2\,\alpha Y(r)\,Y_p(r)\sqrt{Y_{p}(r)}\,Y_1(r)\big]}{56\pi\,(1+Ar^2)^2\,(1+\chi+ \alpha\, Ar^2)^2\,Y(r)} -\frac{8A (1 + Ar^2) (3 + Ar^2)}{56\pi (1 + Ar^2)^3}\nonumber\\&&\hspace{10.2cm} +\frac{4\,A^2r^2[Y_5(r)+Y_6(r)]}{56\pi(1+Ar^2)^3\,Y_p(r)},\label{50}~~~\\
&&\tilde{\rho}=\frac{4A\,\big[4(3-AR^2) + 2 (1 + Ar^2) (3 + Ar^2) (1 + \chi + \alpha \,AR^2)^2+Ar^2\big(Y_5(R)+Y_6(R)\big)\,\big]}{56\,\pi (1 + Ar^2)^3\,(1 + \chi + \alpha \,AR^2)^2}.~~~~~~~~\label{51}
\end{eqnarray}

where,~~$Y_5(r)= (1 + \chi)^2 (5 + Ar^2)+\alpha^2(7 - 15 Ar^2 - 2 A^2r^4),\\
Y_6(r)= 2\alpha (1 + \chi) (-6 + 7 Ar^2 + A^2r^4)$.

\subsection{{Solution of field Equations (\ref{one}) - (\ref{three}) for the source function $\theta_{\mu\nu}$ :}}
In order to determine the components of $\theta_{\mu\nu}$ we need to solve the field equations (\ref{one}) - (\ref{three}). For this purpose we choose the unknown deformation function $f(r)$ to construct a physical viable anisotropic model as,
\begin{eqnarray}\label{decoupler}
f(r)=\frac{Ar^2}{7\,(1+Ar^2)^2}.
\end{eqnarray}
At this point some comments are in order. First, in this opportunity we have imposed a suitable expression for the decoupler function $f(r)$. As should be noted our election (\ref{decoupler}) is a dimensionless expression, free from mathematical singularities and null at the center of the structure \i.e, $f(0)=0$, which implies that $e^{\lambda}|_{r=0}=1$ as required. It is worth mentioning that this methodology has been used in \cite{Ovalle12,Ovalle26}. However, there is another way to get the decoupler function $f(r)$ satisfying the above criteria, in order to close the system of equations (\ref{one})-(\ref{three}). The form to do that is by imposing the so called mimic constraint approach \cite{Ovalle2}. In few words this scheme works by enforcing a direct relation between the seed energy-momentum components and the components of the new source $\theta_{\mu\nu}$ yielding to an algebraic/differential equation allowing to obtain $f(r)$. In this respect both methodologies are completely valid, since we need to prescribe extra information in order to close the problem.

So, by inserting the value of above expression of $f(r)$ into the equations (\ref{one}) - (\ref{three}) we obtain expressions for the $\theta_0^0$, $\theta_1^1$ and $\theta_2^2$ as
\begin{eqnarray}
\theta^0_{0}&=&\frac{A (-3 + A r^2)}{56\,\pi\,(1 + A r^2)^3},\label{53}\\
\theta^1_{1}&=&\frac{-A}{56\,\pi\,(1+Ar^2)}\bigg[1+\frac{8\alpha\,Ar^2}{\sqrt{Y_{p}(r)}}+\frac{4\,D(1+Ar^2) Ar^2}{Y(r)Y_{p}(r)Y_1(r)}\bigg],~~~~~\label{54}\\
\theta^2_{2}&=&\frac{-4\,A^2r^2}{56\,\pi(1+Ar^2)^2}\bigg[\frac{\bar{Y}(r)}{Y(r)}\,Y_7(r)+\frac{2\alpha Ar^2\,Y_8(r)}{(1+Ar^2)\,Y_p(r)\,Y^2_1(r)}\bigg]-\frac{A\,(1-Ar^2)}{56\pi\,(1+Ar^2)^3}\bigg[2Ar^2\frac{\bar{Y}(r)}{Y(r)}+1\bigg].~~~~~~~\label{55}
\end{eqnarray}

where, \\
$\bar{Y}(r)=\bigg[\frac{2\,\alpha\,Y}{\sqrt{Y_p(r)}}+\frac{D\,(1+Ar^2)}{Y_p(r)\,Y_1(r)}\bigg]$,~~~~
$Y_7(r)=\bigg[1+\frac{4\,Ar^2\,(Ar^2-3)}{(1+Ar^2)(A^2r^4+10Ar^2-7)}\bigg]$,\\
$Y_8(r)=4\,(A^2r^4-2Ar^2-3)-\alpha\,(A^3r^6+7A^2r^4+15Ar^2-7)$.

\subsection{General solution of Einstein-Maxwell field equations for Minimal Geometric Deformation (MGD) decoupling source $\theta_{\mu\nu}$:}

The complete solution of Einstein-Maxwell field equations for MGD decoupling source $\theta_{\mu\nu}$ is given as,

\begin{eqnarray}
ds^2=Y^2\,dt^2-\bigg[\frac{7-(10-\alpha)\,A\,r^2-A^2\,r^4}{7 + 14 Ar^2 + 7A^2\,r^4}\bigg]^{-1}dr^2-r^2(d\theta^2+\sin^2\theta\,d\phi^2).\label{dsMGD}
\end{eqnarray}

where $Y$ is given by Eq.(\ref{Sol1}). Now the effective density ($\rho^{eff}$), effective radial pressure ($p_r^{eff}$) and effective tangential pressure ($p_t^{eff}$) can be determined by using Eqs.(\ref{effecrho}-\ref{effecpt}) and (\ref{53}-\ref{55}) as
\begin{eqnarray}
\rho^{eff}&=&\tilde{\rho}+\frac{\beta\,A (-3 + A r^2)^2}{56\,\pi\,(1 + A r^2)^3},\\
p_r^{eff}&=&\tilde{p}+\frac{A\,\beta}{56\,\pi}\bigg[\frac{\sqrt{Y_{p}(r)}+8\alpha\,Ar^2}{\sqrt{Y_{p}(r)}\,(1+Ar^2)}+\frac{4\,D Ar^2}{Y(r)Y_{p}(r)Y_1(r)}\bigg],\\
p_t^{eff}&=&\tilde{p}-\resizebox{0.85\hsize}{!}{$\frac{A\,(1-Ar^2)}{56\pi\,(1+Ar^2)^3}\bigg[\frac{2Ar^2\,\bar{Y}(r)+Y(r)}{Y(r)}\bigg]+\frac{4\,\beta\,A^2r^2}{56\,\pi\,\zeta(r)}\bigg[\,\frac{\bar{Y}(r)\,Y_7(r)}{Y(r)}+\frac{2\alpha Ar^2\,Y_8(r)}{(1+Ar^2)\,Y_p(r)\,Y^2_1(r)}\bigg]$}.~~~~~~~~
\end{eqnarray}

where $\tilde{\rho}$ and $\tilde{p}$ are given in Eqs. (\ref{50})-(\ref{51}) and \\$\zeta(r)=(1+Ar^2)^2$.
\section{Matching conditions}\label{sec5}
In the study of stellar interiors a crucial and important point  is joint  the interior space time geometry with the exterior in a smooth way. To so do, we will employ the well known Israel-Darmois junction conditions ~\cite{r78,r79}. These condition require the continuity of the metric potentials across the surface $\Sigma$ (defined by $r=R$)
\begin{equation}
\left[ds^{2}\right]_{\Sigma}=0,
\end{equation}
or equivalently
\begin{eqnarray}
e^{\nu^{-}(r)}|_{r=R}&=&e^{\nu^{+}(r)}|_{r=R}, \\
e^{\lambda^{-}(r)}|_{r=R}&=&e^{\lambda^{+}(r)}|_{r=R}.
\end{eqnarray}
The above expressions are known as the first fundamental form. On the other hand, a vanishing radial pressure at the boundary of the star is required,
\begin{equation}
p_{r}(r)|_{r=R}=0 \Rightarrow p^{eff}_{r}(r)|_{r=R}=0.
\end{equation}
This condition determines the size of the compact object i.e the radius $R$ and it is known as the second fundamental form. An important point to note here is that these conditions work very well in the context of GR, where the matter distribution remains confined within the compact structure and depending on the ingredients that this material content has, the joint is made with the external Schwarzschild solution (empty outside), Reissner-Nordstr\"{o}m solution (outside with electric charge), etc. Furthermore, in the present case of $f(R,\mathcal{T})$ gravity, the modifications introduced by $\mathcal{T}$ can change slightly or totally the above junction conditions given by (63), (64) and (65). However, the choice of the model f(R, $ \mathcal{T})$ and the Lagrangian density $\mathcal{L}_{m}$ given by (3) ensure that the material content is confined inside the star, so it means ${\mathcal{T}}=0$ beyond the stellar model. This implies that the junction conditions described above can be applied here. Moreover, since the model contains a charge, the continuity of the electric charge $q(r)$ on the surface of the object must be added to the first fundamental form.
\begin{equation}
q(r)|_{r=R}=Q,
\end{equation}
where $Q$ is total object charge. On the other hand, if the electric charge is continuous function of the radial coordinates across the boundary also the electric field $E(r)$ is. Then, the Riemannian exterior manifold can be described by the well known Reissner-Nordstr\"{o}m space time as follows
\begin{equation}
 ds^{2} =\left(1-\frac{2M}{r}+\frac{q^2}{r^2}\right) dt^{2} -\left(1-\frac{2M}{r}+\frac{q^2}{r^2}\right)^{-1}\,dr^{2} -r^{2} \left(d\theta ^{2}+\sin ^{2} \theta  d\phi ^{2}\right).
\end{equation}
So the explicit expression for the first fundamental form as given by
\begin{eqnarray}
 \frac{7-(10-\beta)\,A\,r^2-A^2\,r^4}{7 + 14 Ar^2 + 7A^2\,r^4}&=&\bigg(1-{\frac {2M}{R}}+\frac{Q^2}{R^2}\bigg),\\
  Y^2(R)&=& \left( 1-{\frac {2M}{R}} +\frac{Q^2}{R^2}\right).
 \end{eqnarray}
 where the charge continuity was considered in the electric field continuity. Here the parameter $M$ represents the total mass containing within the star. Solving the second fundamental form given by (65) and the equations obtained from the first fundamental form expressed by (68)-(69) one arrives at the following system of equations for the constant parameters $C$, $D$ and $M$

\begin{eqnarray}
&&\resizebox{0.97\hsize}{!}{$\frac{C}{D}=-\frac{F(R)\, Y_p(R)\,\big[7 \alpha^2\, Y_1^2(R) + (1 + \chi)\, F_3(R) -
   \alpha \left(\sqrt{g_{\chi\alpha}}\, Y_1(R) \,(-7 + 5 AR^2) - 10 (1 + \chi) \, Y_1^2(R)\right)\big]}{Y_1(R)\, \big[(7 \alpha^2 + 10 \alpha (1 + \chi) - (1 + \chi)^2) AR^2 + \sqrt{g_{\chi\alpha}}\, \big( 7\alpha - 5\alpha\, AR^2 + (1 + \chi) \big(5 + AR^2)\big)\big] Y_p(R)}$}\nonumber\\ &&\hspace{8cm}-\resizebox{0.41\hsize}{!}{$\frac{4(1 + AR^22)^2 [-7 + (10-\beta) AR^2 + A^2R^4]}{\big[F_1(R)+F_2(R)\big]\, Y_1(R)\, Y_p(R)}$},\\
&&D=\frac{1}{Y_p(R)\,\big[\frac{C}{D}+\,F(R)\big]}\,\sqrt{\frac{7-(10-\beta)AR^2-A^2R^4}{7+14AR^2+7A^2R^4}}, \\
&& \resizebox{1\hsize}{!}{$M=\frac{R}{2}\,\Bigg[\frac{AR^2(24-\beta+8AR^2)}{7\,(1+AR^2)^2}+\frac{4A^2R^2\,\big[(1 +\chi)^2\,(5 + AR^2) + \alpha^2\,(7 -15AR^2 - 2A^2R^4) + 2\alpha(1 + \chi)\,M_1(R)\big]}{7(1 + AR^2)^3\,(1 + \chi + \alpha\,AR^2)^2}\Bigg]$}.~~~~~~~~
\end{eqnarray}

where,\\\\
$Y_{p}(R)=(1+\chi+\alpha\,AR^2)^2$,~~$Y_1(R)=\sqrt{7- 10 AR^2 - A^2R^4},~~~M_1(R)=(-6 + 7AR^2 + A^2R^4)$\\\\~~ ${Y_2}(R)=3 (1 + \chi)^5 + 3 \alpha (1 + \chi)^4 (-8 + AR^2) + \alpha^2 (1 + \chi)^3  (185 + 4 AR^2 + A^2R^4)$, \\\\
$Y_3(R)=a^3 (1 + \chi)^2 (-535 + 142 AR^2 + 7 A^2R^4) + 3 a^4 (1 + \chi) (-84 - 316 AR^2 + 21 A^2R^4)-\alpha^5 (49 + 161\,AR^2 + 359\, A^2R^4)$,~~~
$Y_4(R)= [5 + 5\,\chi + a (7 - 5 AR^2) +AR^2] $,\\
$F(R)=\frac{{Y_1(R)}\,\big[{Y_2}(R)+Y_3(R)\big]}{3\,(g_{\chi\,\alpha})^3\, (1 +
\chi + \alpha AR^2)^3}
+\frac{4\,f_{\alpha\chi}}{\big(g_{\chi\alpha}\big)^{7/2}}\,\Large{\ln}\bigg[
\frac{\alpha\,(g_{\chi\alpha})^3\,\left[Y_{4}(R) + \chi
AR^2\sqrt{g_{\chi\alpha}}\,\,{Y_1(R)}\right]}{2\,\sqrt{g_{\chi\alpha}}\,f_{
\alpha\chi}\,\,(1 + \chi + \alpha AR^2)}\bigg]$, \\\\
$ F_1(R)= \alpha^2 AR^2 \big[-84+ 9 (12 - \beta) AR^2 + (128 - 9 \beta) A^2R^4 + 16 A^2R^3] - (1 + \chi)^2 [\beta (1 + AR^2) - 4 (6 + 3 AR^2 + A^2R^4)\big] $,\\
$ F_2(R)=2 a (1 + \chi)\, [-28 - 5 (-12 + \beta) AR^2 + (48 - 5 \beta) A^2R^4 + 8 A^3R^6] $,\\
$F_3(R)={Y_1(R)\, \big[(7 \alpha^2 + 10 \alpha (1 + \chi) - (1 + \chi)^2) AR^2 + \sqrt{g_{\chi\alpha}}\, \big( 7\alpha - 5\alpha\, AR^2 + (1 + \chi) \big(5 + AR^2)\big)\big] Y_p(R)}$. \\

 This system of equations contains all the necessary information to find the constant and physical parameters of the present model.
\section{Main salient features}\label{sec6}
\begin{figure}
\includegraphics[width=7.5cm]{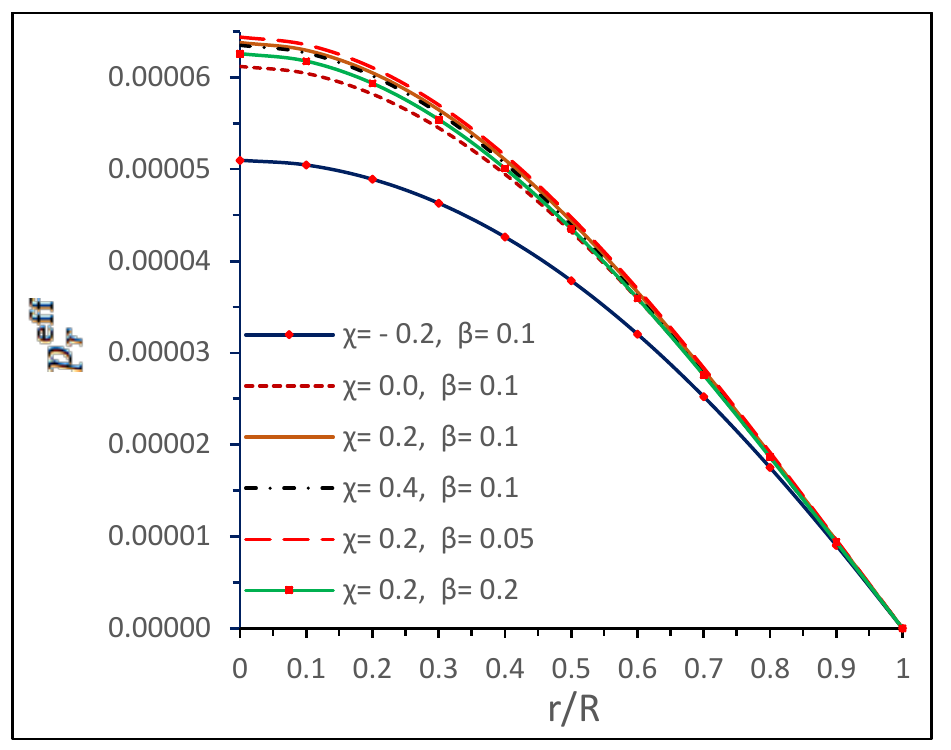}\includegraphics[width=7.5cm]{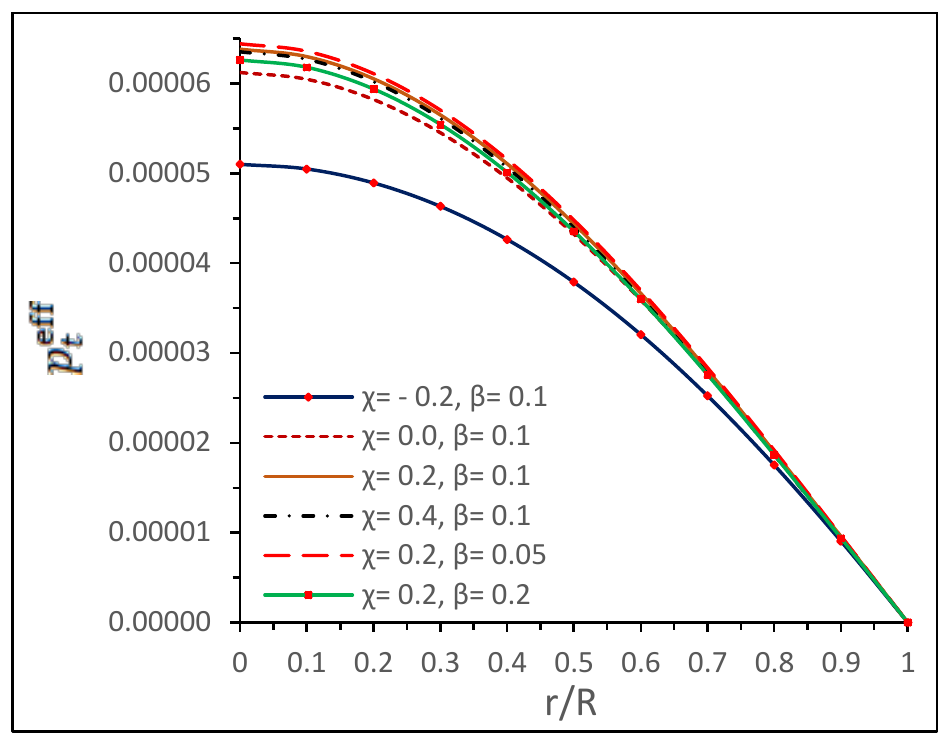}
\caption{\label{fig1} Behavior of effective radial pressure $p_r^{eff}$ and effective tangential pressure $p_t^{eff}$ ves. radial coordinates $(r/R)$.}
\end{figure}

\begin{figure}
\includegraphics[width=7.5cm]{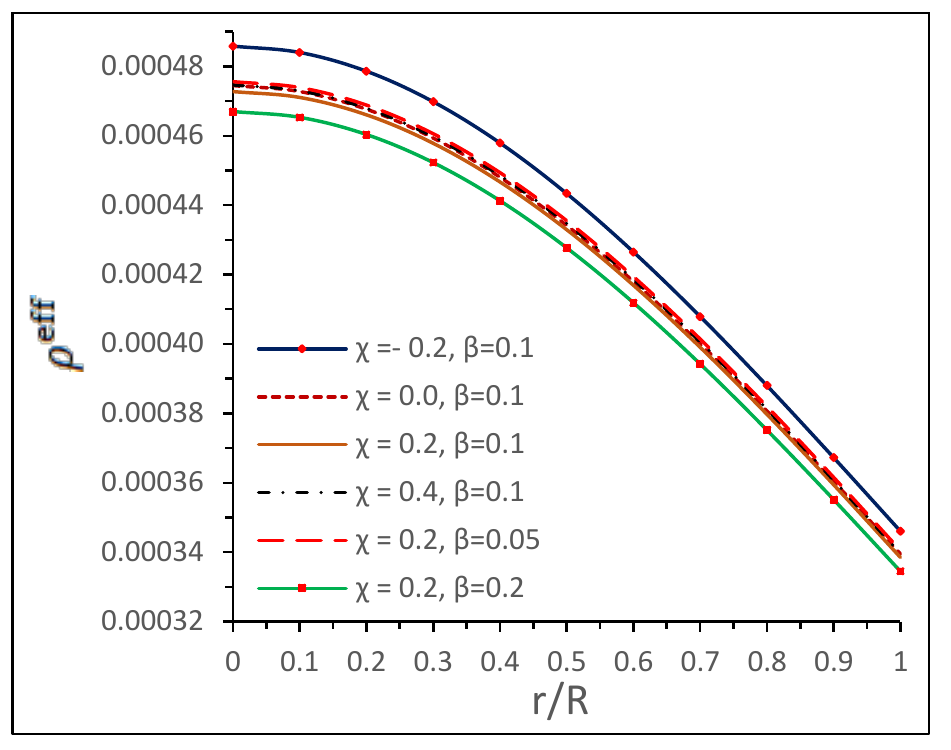}\includegraphics[width=7.5cm]{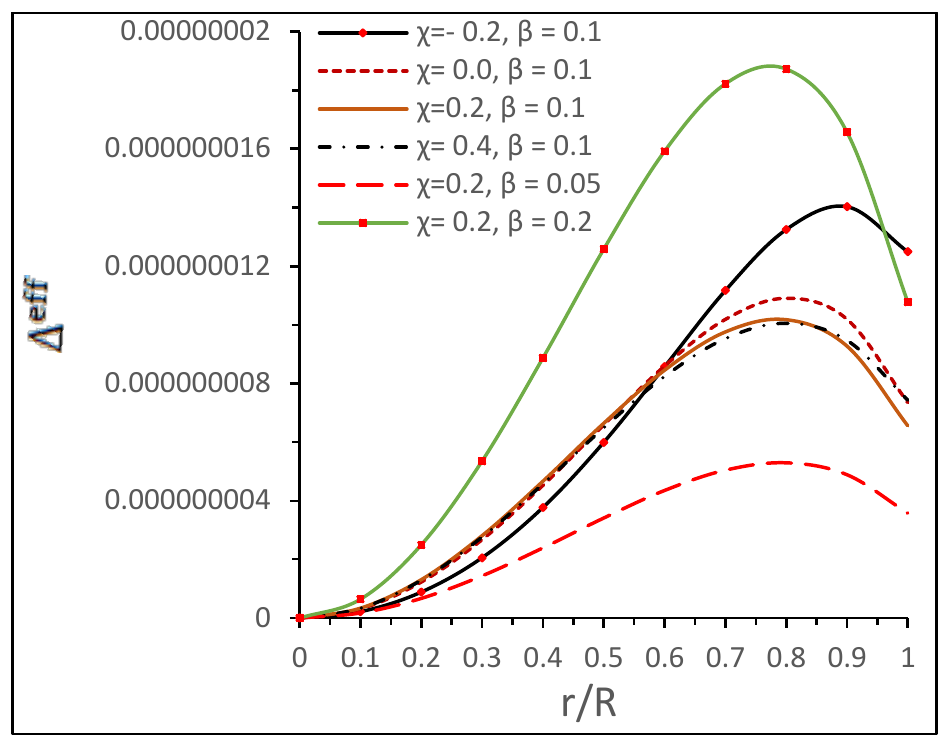}
\caption{\label{fig2} Behavior of effective density  $\rho^{eff}$ and effective anisotropy  $\Delta^{eff}$ves. radial coordinates $(r/R)$.}
\end{figure}

\begin{figure}
\centering
\includegraphics[width=7.5cm]{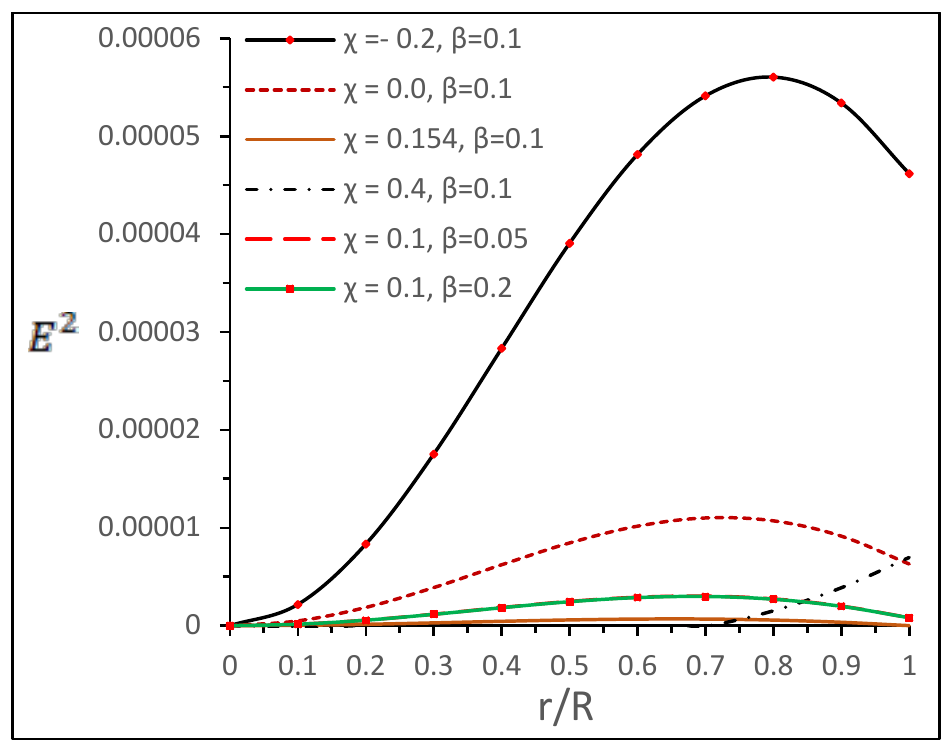}
\caption{\label{fig3} Behavior of effective electric field  $E^2$ ves. radial coordinates $(r/R)$.}
\end{figure}

In this section we presented the main physical quantities that characterize the system. These quantities are the effective matter density $\rho^{eff}$, the effective radial $p^{eff}_{r}$ and tangential $p^{eff}_{t}$ pressure, electric field $E$ and anisotropy factor $\Delta$. It is well known that in the study of charged anisotropic compact structures all the quantities mentioned above must satisfy some general requirements in order to describe a well behaved stellar interior. These general conditions are
\begin{itemize}
    \item Non physical and mathematical singularities at every point within the object exists.
    \item All the thermodynamic observables such as $\rho^{eff}$, $p^{eff}_{r}$ and $p^{eff}_{t}$ must have their maximum values at the center of the configuration which implies that they are monotonic decreasing function with increasing radial coordinate towards the boundary of the star.
    \item Respect to the anisotropy factor $\Delta$, it must be zero at the center, this is so because at the center of the star $p^{eff}_{r}=p^{eff}_{t}$. On the other hand its values should be positive towards the surface (in the case of repulsive anisotropic force).
    \item The electric field $E$ must attains its maximum value at the surface and must be completely null at the center.
\end{itemize}
The detailed behaviour of these quantities is shown in Figs. \ref{fig1}, \ref{fig2} and \ref{fig3}. In these plots one can see that all the physical quantities that describe the system fulfill the general requirements. Some comments regarding the physical and mathematical behaviour of these observables are given below
\begin{itemize}
    \item Figs.\ref{fig1} exhibit the behaviour of the effective radial $p^{eff}_{r}$ and tangential $p^{eff}_{t}$ pressure respectively. Both quantities have their maximum values at the center and decrease monotonically towards the boundary with increasing radius. It is observed that $p^{eff}_{r}$ is vanished at the surface, this fact determines the object size \i.e the radius of the star. Furthermore, the value of $\chi$ has a preponderant effect on both pressures, as we can see varying $\chi$ from $-0.2$ to $0.4$ setting $\beta=0.1$ the upper value of $p^{eff}_{r}$ and $p^{eff}_{t}$ increases with increasing $\chi$. However Fig.\ref{fig2} (left panel) displays the effective density $\rho^{eff}$ behaviour. This quantity has its maximum values at the center of the star and decreases monotonically towards the boundary. Furthermore, is positive defined everywhere within the object.
   \item Fig. \ref{fig2} (right panel) shows the behaviour of anisotropy factor $\Delta$ everywhere inside the star. At the center of the star $\Delta=0$, in effect at the center $p^{eff}_{r}=p^{eff}_{t}$. Moreover, as $\beta$ increase $\Delta$ grow up. Additionally, the anisotropy factor is positive $\Delta>0$, it means that $p^{eff}_{t}>p^{eff}_{r}$ and therefore one has a more compact and massive structures. In this case, the system experiences a repulsive force that counteracts the gravitational gradient improving the equilibrium condition and stability.
   \item Fig.\ref{fig3} displays the squared electric field behaviour. For fixed values of $\beta=0.1$ and $\alpha=1.177$, The squared electric field $E^2$ gradually decreases when move $\chi=-0.2$ to $\chi=0.4$. It is remarkable that $E^2$ have negative values within the stellar model when $0.154 < \chi < 0.643$ whilst at surface of stellar model remains positive for all values of $\chi$ and completely null at the center.
\end{itemize}

%%%%%%%%%%%%%%%%%%%%%%%%%%%%%%%%%%%

%%%%%%%%%%%%%%%%%%%%%%%%%%%%%

\section{Dynamical Equilibrium and Stability Condition}\label{sec7}
 \subsection{Modified Tolman-Oppenheimer-Volkoff (TOV) equation under  Minimal Geometric Deformation (MGD) decoupling source $\theta^{\mu}_{\nu}$ in the framework of $f\left(R,\mathcal{T}\right)$ theory gravity}\label{5.21}

\begin{figure}[htp!]
\centering
\includegraphics[width=7.5cm]{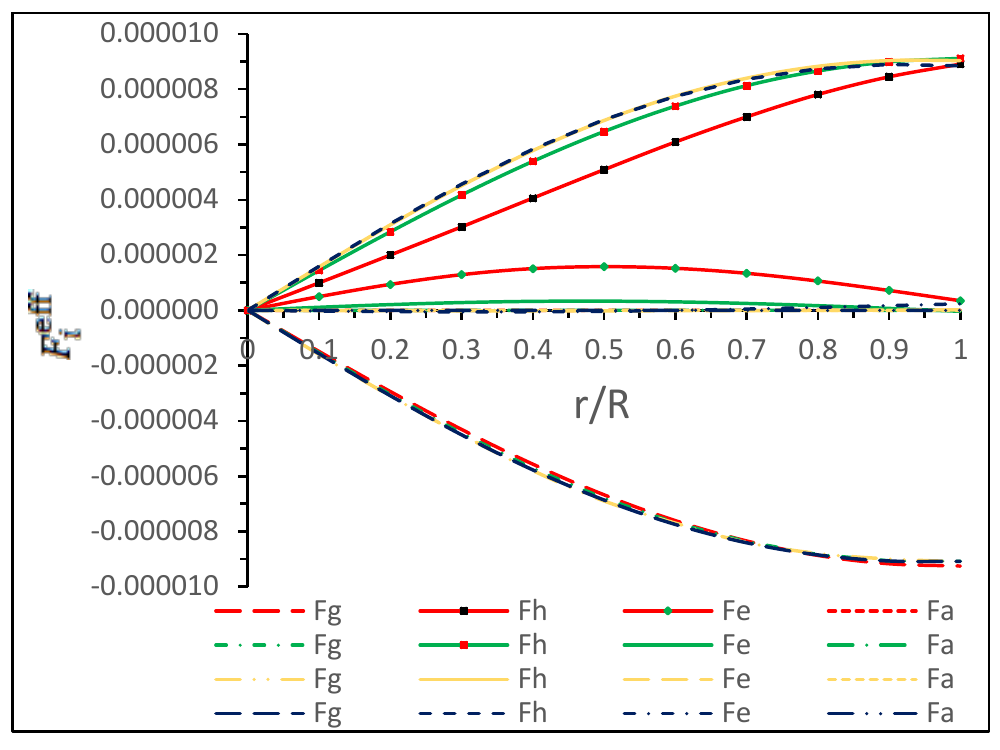}\includegraphics[width=7.3cm]{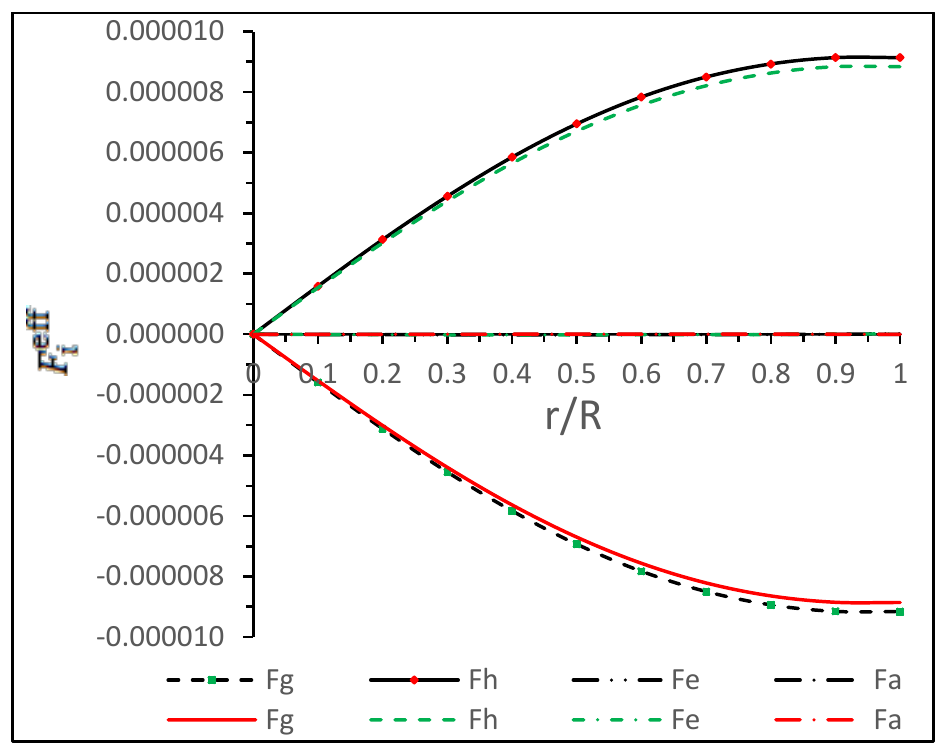}
\caption{\label{fig6}Behavior of different effective forces ves. radial coordinates $(r/R)$. For plotting of thid Fig. we have employed the following set of values: For 1st (left panel) graph: $\chi=-0.2, \alpha=1.177, \beta=0.1$ (red lines), $\chi=0.0, \alpha=1.177, \beta=0.1$ (green lines), $\chi=0.2, \alpha=1.177, \beta=0.1$ (yellow lines), $\chi=0.4, \alpha=1.177, \beta=0.1$ (black lines), For 2nd (right panel) graph: $\chi=0.2, \alpha=1.177, \beta=0.05$ (black lines), $\chi=0.2, \alpha=1.177, \beta=0.2$ (red and green lines). }
\end{figure}

In this section we consider the modified TOV ~\cite{r80,r81} equation in the framework of $f\left(R,\mathcal{T}\right)$ gravity theory for Minimal Geometric Deformation (MGD) decoupling source $\theta^{\mu}_{\nu}$ to perform the equilibrium analysis of the model. It is well known that TOV equation come from the conservation of the energy-momentum tensor (Bianchi's identity) \i.e
\begin{equation}
\nabla^{\mu}T_{\mu\nu}=0.
\end{equation}
As we explained before the election of the function  $f\left(R,\mathcal{T}\right)$ and Lagrangian matter $\mathcal{L}_{m}$ lead us to the conservation of the effective energy-momentum tensor (22) which finally is expressed as follows
\begin{equation}
\nabla^{\mu}{T}^{eff}_{\mu\nu}=0.
\end{equation}
Explicitly,
\begin{eqnarray}\label{tov}
-\big(\tilde{p}-\beta\,\theta^{1}_{1}\big)^{\prime} -\frac{\nu^{\prime}}{2}\,(\tilde{\rho}+\tilde{p}+\beta\,\theta^{0}_{0}-\beta\,\theta^{1}_{1})+\frac{2\,q\,q^{\prime}}{8\,\pi\,r^4}-\frac{2\,\beta}{r}\left(\theta^{2}_{2}-\theta^{1}_{1}\right)=0. ~~~
\end{eqnarray}
As we can see, equation (\ref{tov}) reflexes that the system is under four forces

\begin{eqnarray}
 \textrm{Effective hydrostatic force}:
F_h^{eff}&=&\big(\frac{d\tilde{p}}{dr}-\beta\,\frac{d\theta^{1}_{1}}{dr}\big),
\\
  \textrm{Effective gravitational force}: F_g^{eff}&=&-\frac{\nu^{\prime}}{2}\,(\tilde{\rho}+\tilde{p}+\beta\,\theta^{0}_{0}-\beta\,\theta^{1}_{1}),\\
  \textrm{Effective electric force}: F_e^{eff}&=&\frac{2\,q\,q^{\prime}}{8\,\pi\,r^4}\\
  \textrm{Effective anisotropic force}: F_a^{eff}&=&-\frac{2\,\beta}{r}\left(\theta^{2}_{2}-\theta^{1}_{1}\right).~~~~~~~
 \end{eqnarray}
Therefore, these forces establish the equilibrium of the system. The behavior of these forces is strongly linked to the values taken by the constants $\beta$, $\alpha$ and $\chi$. From Fig. \ref{fig6} we can see how the equilibrium of the system is reached under these different forces. So, we can highlight several points according to the different values assigned to the parameters $\beta$, $\alpha$ and $\chi$ in Fig. \ref{fig6}.
%%%%%%%%%%%%%%%%%%%%%%
%%%%%%%%%%%%%%%%%%%%%
\begin{itemize}
    \item Left panel: In this plot, $\alpha$ and $\beta $ parameters take values $ 1.177$ and $0.1$ respectively, while $\chi$ parameter takes values from $-0.2$ to $0.4$. It is noteworthy the effect of $\chi$ parameter has on the behavior of hydrostatic, electric and gravitational forces. As we can see, when $\chi$ increases the hydrostatic and gravitational forces (in negative direction) increase but the electric one decreases. Although the electric force decreases when $\chi$ increases, the increment of the hydrostatic force in this direction helps to maintain the equilibrium of the system (the same happens when $\chi$ decreases, in this case the electric force increases and the hydrostatic force decreases), this means that there is a complement between the behavior of both forces to counteract the gravitational gradient. On the other hand, the anisotropic force, despite being very small in comparison to the other forces, is repulsive in nature since $\Delta> 0$, which helps to counterbalance the gravitational force together with the hydrostatic and electric forces.
    \item Right panel: In distinction with the left panel here $\chi$ and $\alpha$ are fixed to $0.2$ and $1.177$ respectively, till $\beta$ varies from $0.05$ to $0.2$. In this case both the anisotropic and the electric forces are small in comparison with the hydrostatic and gravitational ones. This time, the gravitational and hydrostatic gradients decrease in a proportional way when $\beta$ decreases (both gradients increase otherwise). Nevertheless, the system remains in equilibrium.    \end{itemize}
 According to the previous discussion, it is clear what is the response of the different forces to which the compact object is subjected to remain in equilibrium, under the different values that take the coupling constants $\alpha$, $\beta$ and $\chi$.

 \subsection{Measure of stability for the stellar compact objects via cracking}\label{5.2}
\begin{figure}[htp!]
\includegraphics[width=7.5cm]{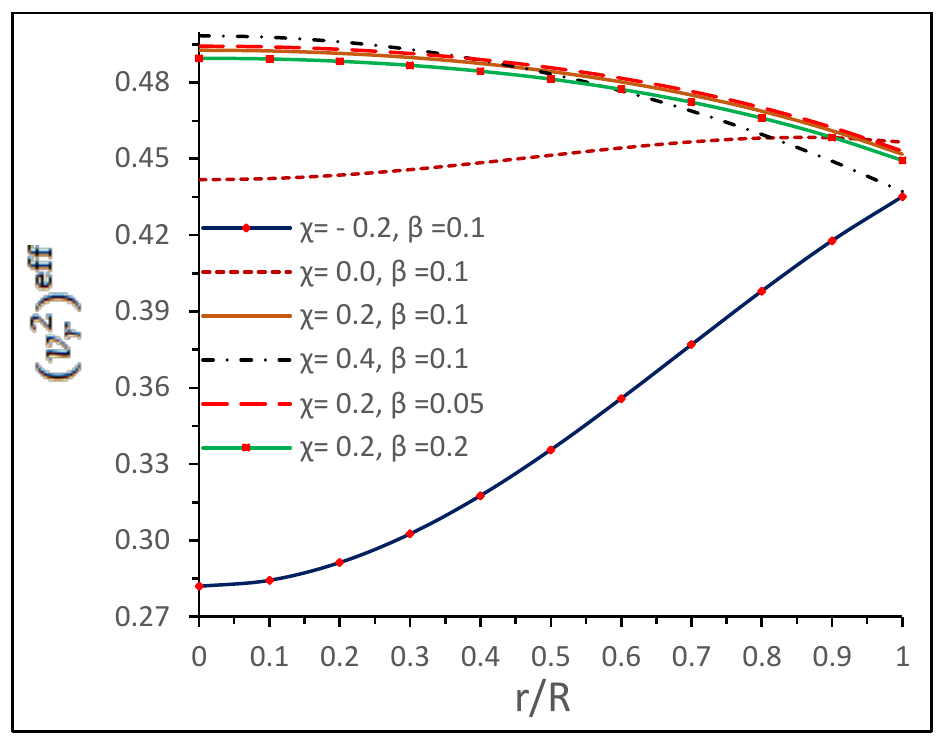}\includegraphics[width=7.5cm]{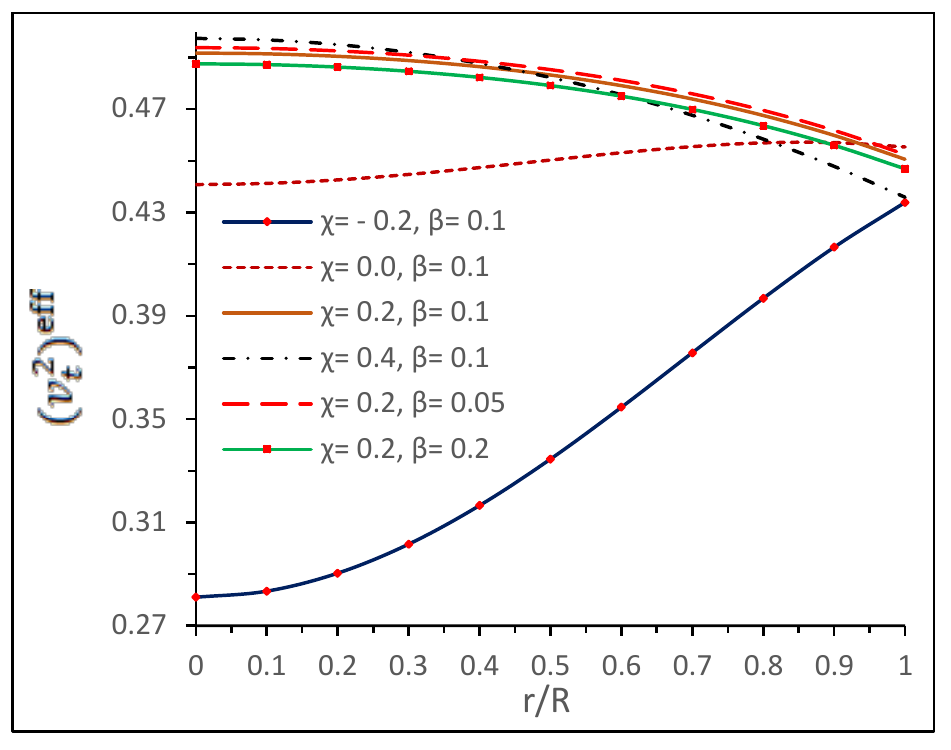}
\caption{\label{fig7}Behavior of effective square radial velocity  $(v^2_r)^{eff}$ and effective square tangential velocity  $(v^2_t)^{eff}$ ves. radial coordinates $(r/R)$.}
\end{figure}

\begin{figure}[htp!]
\includegraphics[width=7.5cm]{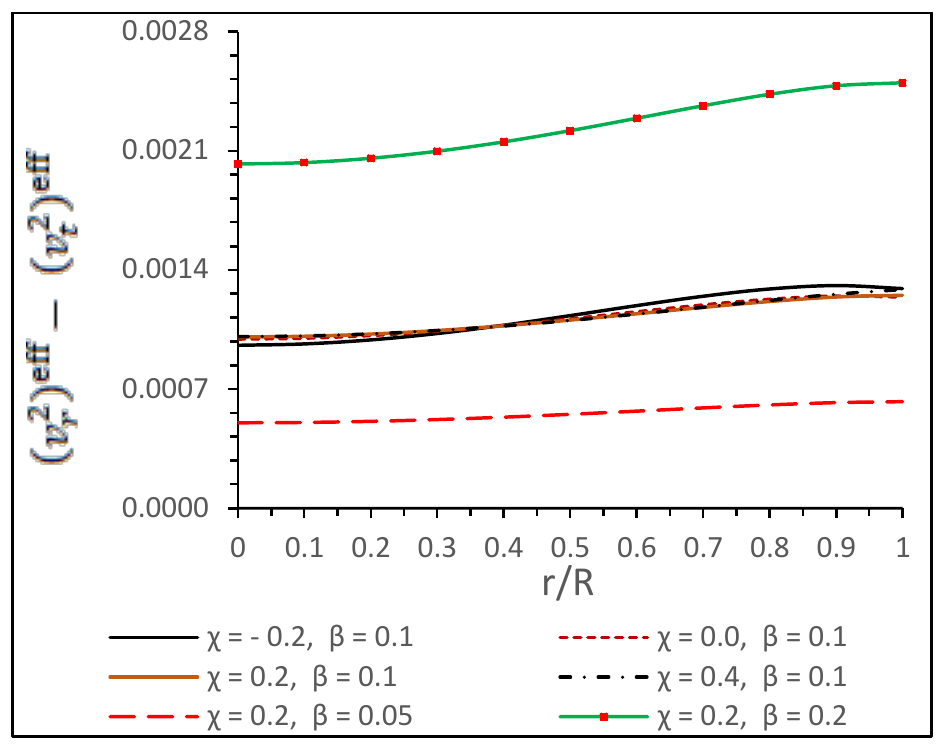}\includegraphics[width=7.5cm]{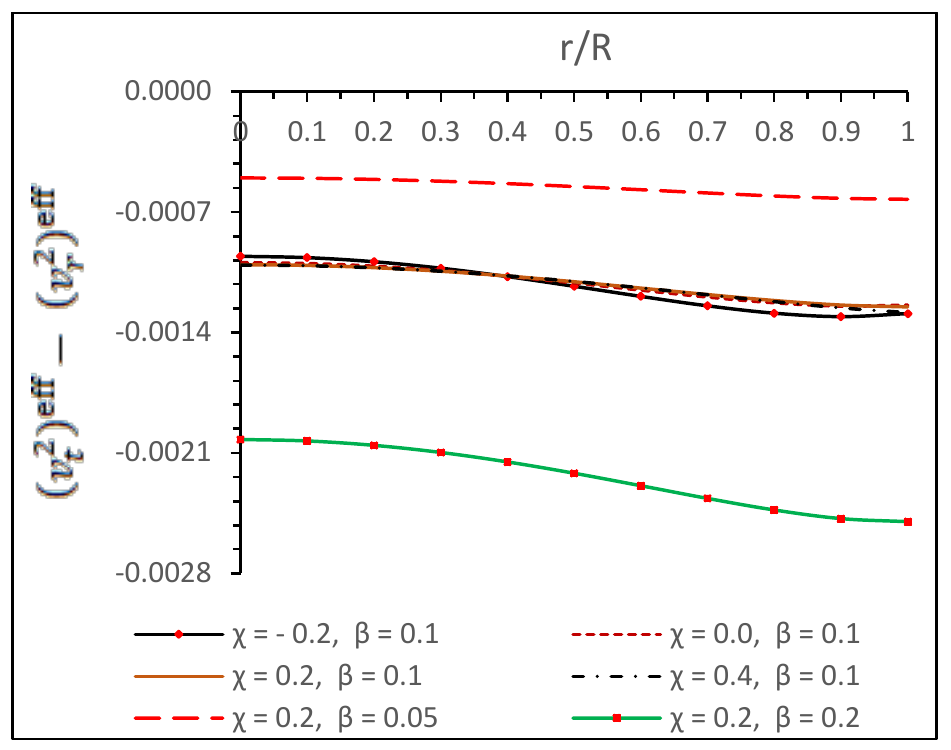}
\caption{\label{fig8}Behavior of difference of effective square radial velocities  $(v^2_r)^{eff}-(v^2_r)^{eff}$ and $(v^2_t)^{eff}-(v^2_r)^{eff}$ ves. radial coordinates $(r/R)$.}
\end{figure}

Once the equilibrium of the system has been analyzed a question arises, Is this equilibrium stable or not?. In this case we need to investigate if the perturbation introduced by the anisotropic source $\theta_{\mu\nu}$ yields to the system a stable equilibrium. To do so, we use Abreu's criteria based on Herrera's cracking concept ~\cite{r82}. Basically, this criterion studies the stability of compact objects through the subliminal radial and tangential sound speeds, it means
\begin{eqnarray}
     \label{169}
   -1\leq v^{2}_{t}-v^{2}_{r}\leq 1  = 
 \left\{ \begin{array}{ll}
		   -1\leq v^{2}_{t}-v^{2}_{r}\leq 0~~ & \mathrm{Potentially\ stable\ }  \\
		 0< v^{2}_{t}-v^{2}_{r}\leq 1 ~~ & \mathrm{Potentially\ unstable}
	       \end{array}
	     \right\},~
   \end{eqnarray}
obviously this treatment needs that both, the radial and tangential subliminal sound speeds within the star fulfill causality condition \i.e $v^{2}_{r}<1$ and $v^{2}_{t}<1$ (assuming the speed of light $c=1$ in relativistic geometrized units). Of course, this model satisfies causality condition in the radial and tangential directions, as shows Fig. \ref{fig7}, where it is observed that for negative values of $\chi$ both velocities are increasing in nature (specifically the increasing nature start at $\chi \le 0.097$) and as the value of $\chi$ increases, both are decreasing functions with their maximum values at the center of the compact configuration. On the other hand, Fig. \ref{fig8} shows that the system is completely stable according to Eq.(\ref{169}).

\begin{table}
\caption{\label{table1}Physical values of compact stellar models for $\chi=0.2$, $\alpha=1.177$, $\beta=0.1$.}

\resizebox{1\hsize}{!}{$\begin{tabular}{cccccc}
  {Compact star}&{Mass}& Radius & Surface redshift&
Mass-radius ratio & $AR^2$\\
Models&$M/M_{\odot}$& $R (Km)$&($Z_s$)&$\frac{M}{R}$ &  \\ \hline
EXO 1785-248 ( $\ddot{\text{O}}$zel \emph{et al.} \cite{ozel}) &  1.3 & 10.5 & 0.25513 & 0.18262 &  0.1311 \\
SMC X-1(Rawls \emph{et al.} \cite{Rawls}) & 1.04 & 9.34 & 0.22031 &  0.16424 & 0.11521 \\
SAX J1808.4-3658 (SS2)(Li \textit{et al.} \cite{Li})& 1.3237 & 6.7 & 0.54739 & 0.29141 & 0.24404\\
Her X-1 ~(Abubekerov \emph{et al.})  \cite{Abubekerov} & 0.85 & 8.1 & 0.20348 & 0.15478 & 0.10732 \\
4U 1538-52 ~(Rawls \emph{et al.} \cite{Rawls}) & 0.87 & 7.866 &0.21833& 0.16314 &  0.11429   \\
LMC X-4(Rawls \emph{et al.} \cite{Rawls})& 1.29 & 10.55 & 0.25067 &  0.18036 & 0.1291\\
SAX J1808.4-3658 ~ (Elebert \textit{et al.} \cite{Elebert}) & 0.9 & 9.5 & 0.17810 & 0.13974 & 0.09515\\
Cen X-3(Rawls \emph{et al.} \cite{Rawls})& 1.49 & 10.8 & 0.298517 & 0.2035 & 0.15011 \\\hline
\end{tabular}$}
\footnotesize\textit{{$^a$In this table I, we have obtained the radius, surface redshift and Mass-radius ratio of the different compact stars for observed mass for $\chi=0.2$, $\alpha=1.177$ and $\beta=0.1$. We have note that the surface redshift is increasing with Mass-radius ratio.}}
\end{table}

%%%%%%%%%%%%%%%%%%%%%%%%%
\begin{table}
\begin{center}
\caption{\label{table2}Comparative study of physical values of compact star LMC X-4  for different values of $AR^2=0.1291$, Mass $ =1.29 M_{\odot}$, $\alpha=1.177$, $\beta=0.1$.}
\resizebox{0.7\hsize}{!}{$\begin{tabular}{ccccc}\hline
 {$\chi$}& Radius  & central pressure &
central density & surface density \\
& $R (Km)$& $(p_r)_c^{eff}$ & $\rho_c^{eff}$ & $\rho_s^{eff}$\\ \hline
-0.2 & 10.40699 &$7.8993\times10^{34}$ & $6.45755\times10^{14}$ &  $4.62303\times10^{14}$ \\
0.0 & 10.5308 &$7.43621\times10^{34}$ & $6.40323\times10^{14}$ &  $4.58222\times10^{14}$ \\
0.2 & 10.55   &$7.75147\times10^{34}$ & $6.37989\times10^{14}$ &  $4.56879\times10^{14}$ \\
0.4 &  10.52854 &$7.71566\times10^{34}$ & $6.405974\times10^{14}$ &  $4.58383\times10^{14}$\\\hline
\end{tabular}$}
\end{center}
\footnotesize\textit{{$^b$In table 2,  we have calculated the radius (R), effective central density ($\rho^{eff}_c$) and effective surface density ($\rho^{eff}_s$), effective central pressure ($p^{eff}_c$) for different values of $\chi$ for the star LMC X-4 with fixed values of $\chi=0.2$ and $\alpha=1.177$. We note that $R$, $\rho^{eff}_c$, $\rho^{eff}_s$ and $p^{eff}_c$ increases from $\chi=-0.2$ to $\chi\approx 0.19$ and then they start decreasing after $\chi\approx 0.19$}}.
\end{table}
%%%%%%%%%%%%%%%%%%%%%%%%%%

\begin{table}
\begin{center}
\caption{\label{table3}Comparative study of physical values of the compact star LMC X-4  for $AR^2=0.1291$ and Mass $=1.29 M_{\odot}$, $\chi=0.2$, $\alpha=1.177$.}
\resizebox{0.7\hsize}{!}{$
\begin{tabular}{ccccc}\hline
 {$\beta$}& Radius  & central pressure & central density & surface density \\
& $R (Km)$& $(p_r)_c^{eff}$ & $\rho_c^{eff}$ & $\rho_s^{eff}$ \\ \hline
0.05 &  10.52939  &$7.82468\times10^{34}$ & $6.41834\times10^{14}$ &  $4.59564\times10^{14}$ \\
0.10 & 10.55  &$7.75147\times10^{34}$ & $6.37989\times10^{14}$ &  $4.56879\times10^{14}$ \\
0.20 & 10.59259  &$7.60557\times10^{34}$ & $6.30225\times10^{14}$ &   $4.51456\times10^{14}$ \\\hline
\end{tabular}$}
\end{center}
\footnotesize\textit{{$^c$In table 3,  we have calculated the radius, effective central and surface density, effective central pressure for different values of coupling constant $\beta$ for the star LMC X-4 with fixed values of $\chi=0.2$ and $\alpha=1.177$. We have note that predicted radius increases when the coupling constant $\beta$ move from $0.05$ to $0.2$ while the effective central and surface density, effective central pressure have an opposite behavior as these physical parameters decreases when $\beta$ increases from $0.05$ to $0.2$}}. 
\end{table}
%%%%%%%%%%%%%%%%%%%%%%
\begin{table}
\begin{center}
\caption{\label{table4}Comparative study of the compact star LMC X-4  for $AR^2=0.1291$ and Mass $ =1.29 M_{\odot}$, $\chi=0.2$, $\beta=0.1$.}
\resizebox{0.7\hsize}{!}{$
\begin{tabular}{ccccc}\hline
 {$\alpha$}& Radius & central pressure &
central density & surface density \\
& $R (Km)$ & $(p_r)_c^{eff}$ & $\rho_c^{eff}$ & $\rho_s^{eff}$\\ \hline
1.000 &  10.52618  & $7.70797\times10^{34}$ & $ 6.40885\times10^{14}$ &  $4.58555\times10^{14}$ \\
1.177 & 10.55   &$7.75147\times10^{34}$ & $6.37989\times10^{14}$ &  $4.56879\times10^{14}$ \\
1.400 & 10.53261  &$7.46387\times10^{34}$ & $6.40101\times10^{14}$ &  $4.56879\times10^{14}$\\ \hline
\end{tabular}$}
\end{center}
\footnotesize\textit{{$^d$ In table IV,  we have calculated the radius, effective central and surface density, effective central pressure for different values of constant $\alpha$ for the star LMC X-4 with fixed values of $\chi=0.2$ and $\beta=0.1$. We have note that predicted radius having same behavior like Table II. It is note that when $\alpha=1$ then the electric charge will purely depends on constant $\chi$ which implies that this charged isotropic solution reduce into isotropic perfect fluid solution in GR when $\chi=0$}}. 
\end{table}

%%%%%%%%%%%%%%%%%%%%%%%

\section{Mass-Radius Ratio and Surface Redshift}\label{sec8}

\subsection{Gravitational and Effective Mass} 
In the study of compact structures it is very important to analyze the maximum limit of the mass-radius ratio. The most simple case, concerning isotropic matter distributions states that the mass-radius ratio is given by the well known Buchdahl's limit \cite{r83}. This limit establishes that when the energy density of the perfect fluid configuration is non increasing outwards the bound is given by
\begin{equation}\label{eq81}
\frac{2M}{R}\leq\frac{8}{9},
\end{equation}
where $M$ is the total mass of the fluid sphere and $R$ the radius of the star in Schwarzschild coordinates, defined by the location of the the vanishing pressure surface. The above limit (\ref{eq81}) has important implications, for example it tells us that the surface redshift $Z_{s}$ has a maximum value, specifically $Z_{s}=2$ (for perfect fluid distributions). Besides, the boundary of a star occurs just after the apparent horizon in the Schwarzschild metric at $r=2M$. However, when the stellar interior contains electric charge Buchdahl's limit is modified. In this case there are an upper \cite{r84} and lower \cite{r85} bounds, 
\begin{equation}\label{eq82}
\frac{Q^{2}\left(18R^{2}+Q^{2}\right)}{2R^{2}\left(12R^{2}+Q^{2}\right)}\leq \frac{M}{R}\leq \frac{4R^{2}+3Q^{2}+2R\sqrt{R^{2}+3Q^{2}}}{9R^{2}}.
\end{equation}
Moreover, the mass $M$ appearing in the expression (\ref{eq82}) is different from the total mass in (\ref{eq81}). Concretely, the total mass in (\ref{eq81}) represents the mass of an isotropic fluid sphere. This amount is exactly equal to the gravitational and effective mass in the perfect fluid case. Explicitly it reads \begin{equation}\label{gravmass}
M=m(R)=4\pi\int^{R}_{0}r^{2}\rho(r)dr=\frac{R}{2}\left[1-e^{-\lambda(R)}\right].
\end{equation} 
While in (\ref{eq82}) $M$ represents the gravitational mass of the charged star, given by
\begin{equation}\label{gravmasschar}
M=m(R)=4\pi\int^{R}_{0}r^{2}\rho^{eff}(r)dr+\frac{1}{2}\int^{R}_{0}\frac{Q^{2}(r)}{r^{2}}dr+\frac{Q^{2}(R)}{2R}=\frac{R}{2}\left[1-e^{-\lambda(R)}+\frac{Q^{2}(R)}{R^{2}}\right]. 
\end{equation}

As depicted by equations (\ref{gravmass}) and (\ref{gravmasschar}), the gravitational mass is larger in the case of electrically charged structures with respect to their uncharged isotropic counterpart. 
\subsection{Surface gravitational redshift}

As was pointed out before, the mass-radius ratio is related with an important quantity \i.e the surface redshift $Z_s$. It can be calculated as
\begin{equation}\label{zs}
Z_s=e^{\lambda/2}-1=\left(1-2u\right)^{-1/2}-1, \quad \mbox{with} \quad u\equiv\frac{M_{eff}}{R}.
\end{equation}
In the present case the efective mass $M_{eff}$ is given by
\begin{equation}\label{effmass}
M_{eff}=m_{eff}(R)=4\pi\int_{0}^{R}\left({\rho^{eff}(r)}+\frac{E^{2}(r)}{8\pi}\right)r^{2}dr=\frac{R}{2}\left[1-e^{-\lambda(R)}\right],
\end{equation}
which does not coincide with the gravitational mass (\ref{gravmasschar}) of the object as occurs in the isotropic case (\ref{gravmass}). Despite equations (\ref{gravmass}) and (\ref{effmass}) have the same final expression (R.H.S), the meaning is not the same. Of course, as said before the gravitational mass of charged configurations has an extra piece \i.e, the electric field contribution. Clearly, from Eq. (\ref{gravmasschar}) the gravitational mass of charged compact objects is grater than the effective mass (\ref{effmass}) in an amount $Q^{2}/R^{2}$. Then it is obvious that the mass-radius relation takes higher values by using (\ref{gravmasschar}) than (\ref{effmass}).

In what respect to gravitational redshift $Z_{s}$, in the case of isotropic matter distribution the maximum value that it can reach is $Z_s=2$, which is in complete agreement with the Buchdahl's limit $u=M_{eff}/R\leq 4/9$. Then from Eq.(\ref{zs}), we observed that the surface redshift of star cannot be arbitrary large due to Buchdahl's limit. Nevertheless, when anisotropies and electric charge are present into the matter content the above limit can be exceeded. In this direction Ivanov \cite{r55}  shown that for a realistic anisotropic star models the upper bound of $Z_s$ is $5.211$ (this value corresponds to a model without cosmological constant) and can not be overcome. As eq. (\ref{zs}) shows $Z_s$ depends on the $e^{\lambda}$ metric potential. Tables \ref{table1}, \ref{table2}, \ref{table3} and \ref{table4} shown different values of $Z_s$ for different values of the parameters $\alpha$, $\beta$ and $\chi$. Notwithstanding, tables \ref{table2} and \ref{table4} display the same $Z_s$ for the same star for different values of $\chi$ and $\alpha$ respectively, while table \ref{table3} shows different values of $Z_s$ varying the $\beta$ parameter. Of course, $e^{\lambda}$ does not depend neither $\chi$ or $\alpha$, only depends on $\beta$, then variation on $\beta$ introduces a modification over $e^{\lambda}$ and in consequence on the surface redshift $Z_s$. This implies that gravitational decoupling parameter $\beta$ affects the surface redshift as well as the compactness factor $u$. It is important to highlight that the compactness factor $u$ is obtained by using the effective mass (\ref{effmass}). This is so because from the observational point of view the electric charge can not be measure at all, therefore one appreciates only a global density (which includes the electric charge), as illustrated in the integral of equation (\ref{effmass}) $\rho^{global}=\rho^{eff}+E^{2}/8\pi$.  
In tables \ref{tab5}, \ref{tab6} and \ref{tab7} are shown the lower and the upper bound of the mass-radius relation for LMC X-4 considering different values of the couplings $\alpha$, $\beta$ and $\chi$. We have considered both the mass-radius ratio taking the gravitational mass (\ref{gravmasschar}) and the effective  one (\ref{effmass}), as was expected the mass-radius relation the former is greater than the second one. Furthemore, the surface redshift values are in accordance what it is reported in the literature.
\begin{table}[hbt!]
\centering
\caption{\label{tab5}Comparative study of lower bound, Mass-radius ratio, upper bound, compactness $(u=M_{eff}/R)$ and surface redshift of the star for $\alpha=1.177$, $\beta=0.1$ with different values of $\chi$.}
\resizebox{1\hsize}{!}{$\begin{tabular}{cccccc}\hline
&Lower bound & Mass-radius ratio & Upper bound & Compactness & Surface redshift\\
$\chi$ &$\frac{Q^2\,(18R^2+Q^2)}{2R^2\,(12R^2+Q^2)}$&$\frac{M}{R}$&$\frac{4R^2+3Q^2+2R\sqrt{R^2+3Q^2}}{9R^2}$&$ \frac{M_{eff}}{R}$& $Z_s$\\
\hline
-0.2 & 3.7509$\times10^{-3}$ & 0.182846 & 0.66999 & 0.180345 & 0.25067 \\
0.0 & 5.2381$\times10^{-4}$ & 0.180694 &0.66713 & 0.180345&0.25067  \\
0.2 & 1.4637$\times10^{-5}$  & 0.180355 & 0.66668  & 0.180345 &0.25067  \\
0.4 & 5.7899$\times10^{-4}$  &  0.180731 & 0.66718  &0.180345 & 0.25067 \\\hline
\end{tabular}$}
\end{table}

%%%%%%%%%%%%%%%%%%%%%%%%%%%%%%%%%%%%
\begin{table}[hbt!]
\centering
\caption{\label{tab6}Comparative study of lower bound, Mass-radius ratio, upper bound, compactness $(u=M_{eff}/R)$ and surface redshift of the star for $\alpha=1.177$, $\chi=0.2$ with different values of $\beta$.}
\resizebox{1\hsize}{!}{$\begin{tabular}{cccccc}\hline
&Lower bound & Mass-radius ratio & Upper bound & Compactness & Surface redshift\\
$\beta$ &$\frac{Q^2\,(18R^2+Q^2)}{2R^2\,(12R^2+Q^2)}$&$\frac{M}{R}$&$\frac{4R^2+3Q^2+2R\sqrt{R^2+3Q^2}}{9R^2}$&$ \frac{M_{eff}}{R}$& $Z_s$\\
\hline
0.05 & 1.46375$\times10^{-5}$ & 0.180717  & 0.66668 & 0.18071&0.25138\\
0.1 & 1.4637$\times10^{-5}$  & 0.180355 & 0.66668 &0.180345&0.25067 \\
0.2 & 1.46375$\times10^{-5}$  & 0.179632 & 0.66668 &0.17962 & 0.24926\\\hline
\end{tabular}$}
\end{table}
%%%%%%%%%%%%%%%%%%%%%%%%%%%%%%%%%%%%%%%

\begin{table}[hbt!]
\caption{\label{tab7}Comparative study of lower bound, Mass-radius ratio, upper bound, compactness $(u=M_{eff}/R)$ and surface redshift of the star for $\beta=0.1$, $\chi=0.2$ with different values of $\alpha$.}
\resizebox{1\hsize}{!}{$\begin{tabular}{cccccc}\hline
&Lower bound& Mass-radius ratio & Upper bound& Compactness & Surface redshift\\
$\alpha$ &$\frac{Q^2\,(18R^2+Q^2)}{2R^2\,(12R^2+Q^2)}$&$\frac{M}{R}$&$\frac{4R^2+3Q^2+2R\sqrt{R^2+3Q^2}}{9R^2}$&$ \frac{M_{eff}}{R}$& $Z_s$\\
\hline
1.0 & 6.31888 $\times10^{-4}$ & 0.180766  & 0.66723&0.180345&0.25067 \\
1.177 & 1.46375$\times10^{-5}$  &  0.180355 & 0.66668 &0.180345&0.25067 \\
1.4 & 4.61906$\times10^{-4}$  & 0.180653 & 0.66708 & 0.180345&0.25067  \\\hline
\end{tabular}$}
\end{table}

%%%%%%%%%%%%%%%%%%%%%%%%%%%%%%%%%%

\section{Energy conditions}\label{sec9}
\begin{figure}[hbt!]
\includegraphics[width=4cm]{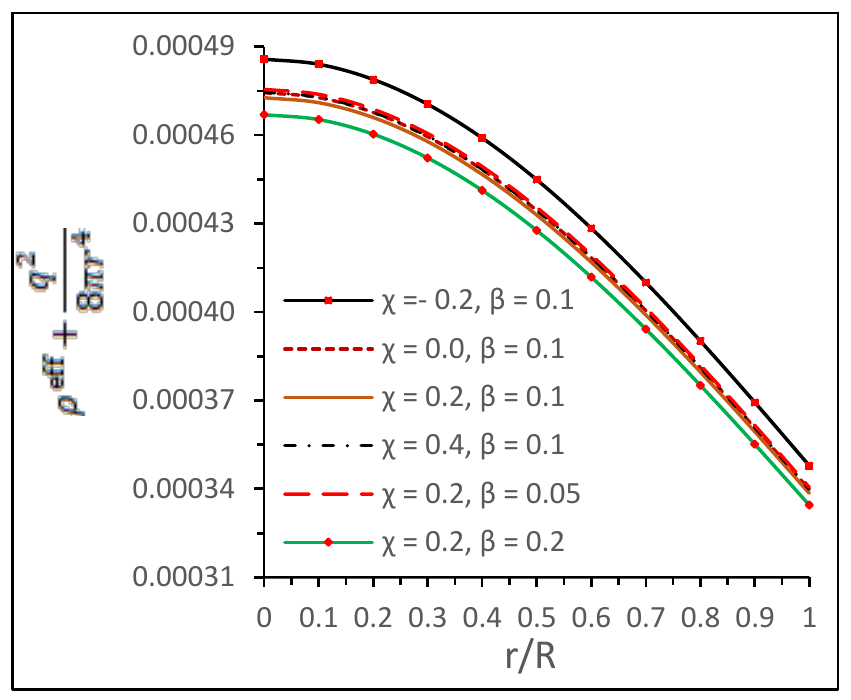}\includegraphics[width=4cm]{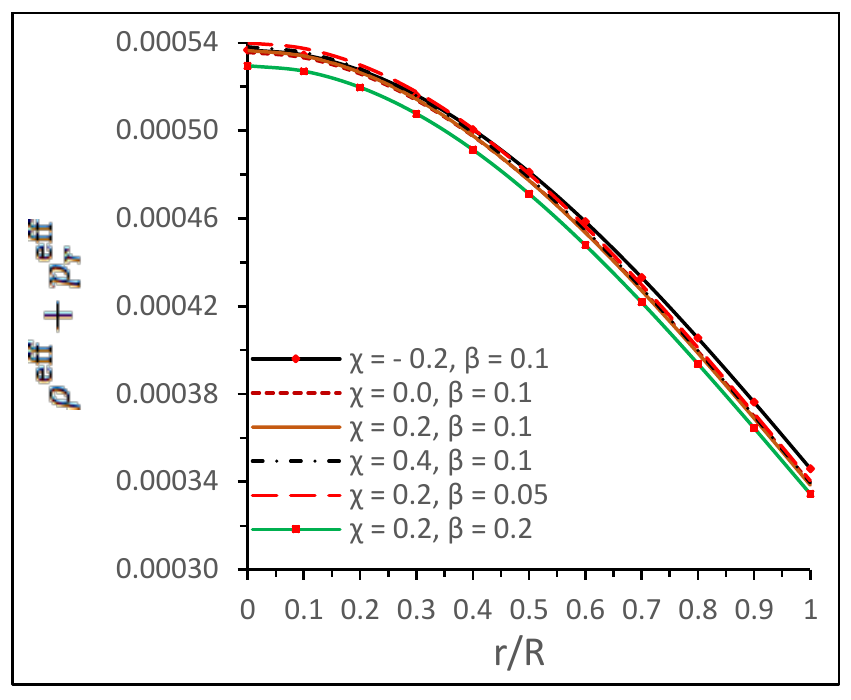}\includegraphics[width=4cm]{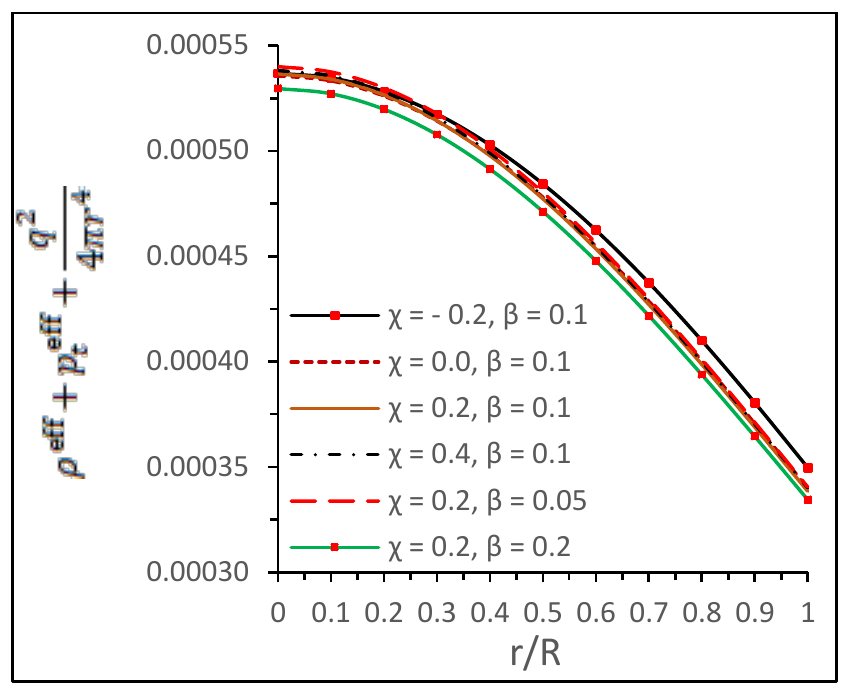}\includegraphics[width=4cm]{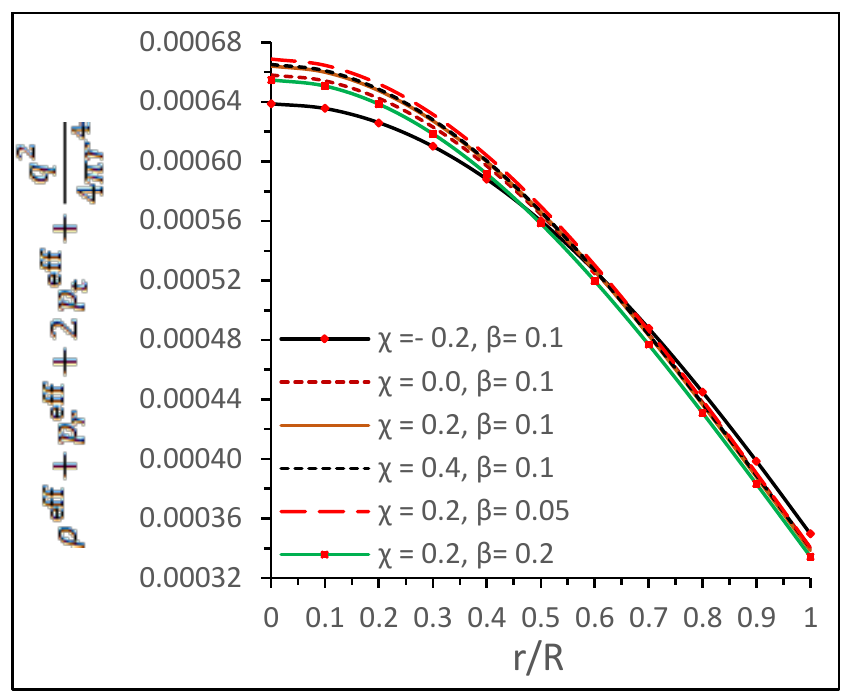}\\\includegraphics[width=4cm]{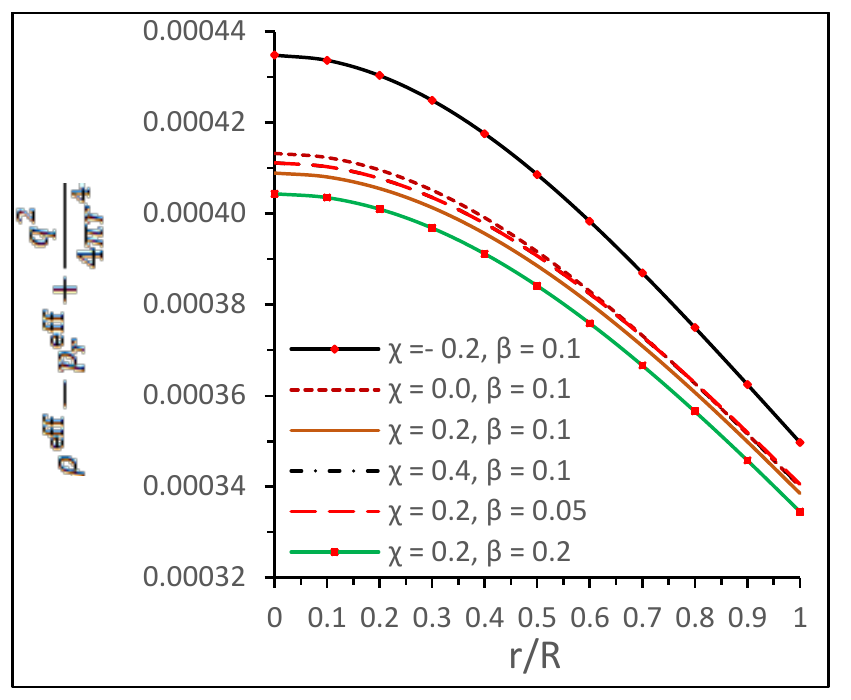}\includegraphics[width=4cm]{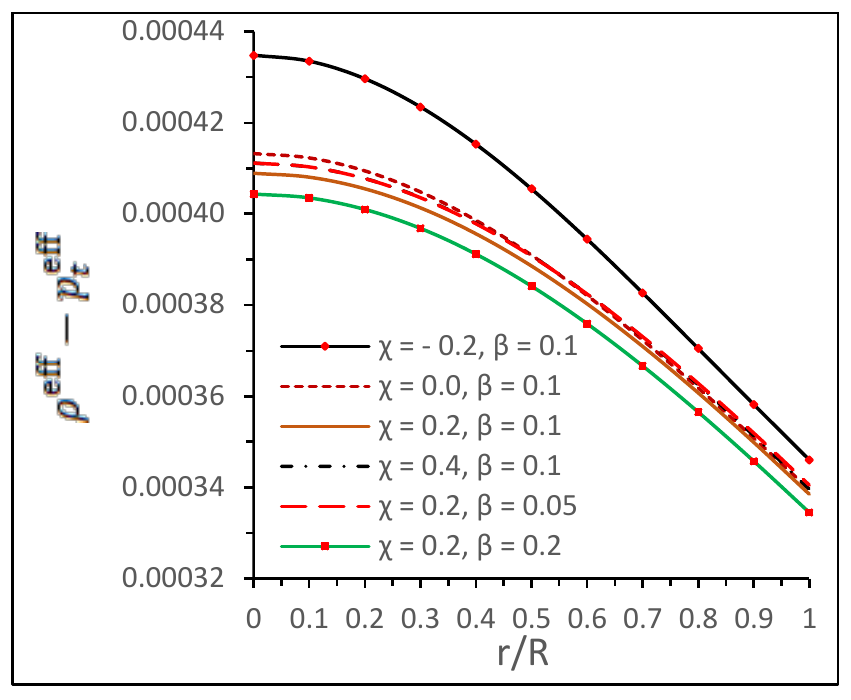}\includegraphics[width=4cm]{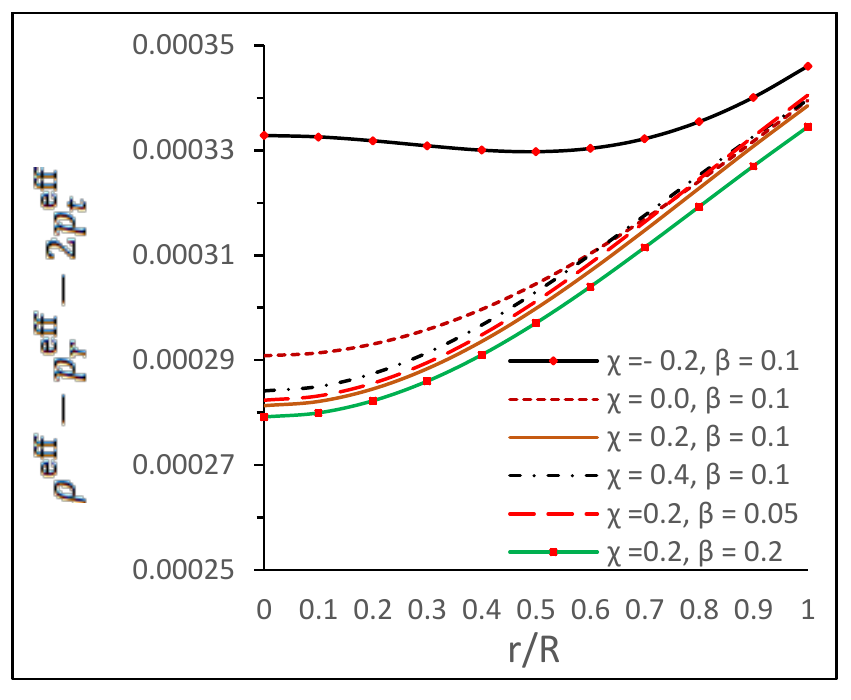}
\caption{\label{fig9}Behavior of energy conditions ves. radial coordinates $(r/R)$.}
\end{figure}

Independent of the framework used to study collapsed structures, one of the most important thing concern a positive and well defined energy-momentum tensor everywhere inside the star. To check the above requirements the matter content must satisfies the following inequalities simultaneously
\begin{eqnarray}
NEC &:&  \rho^{eff}+p_r^{eff} \ge 0\\
 WEC &:&   \rho^{eff}+\frac{q^2}{8\pi\,r^4}\ge 0,~\rho^{eff}+p_r^{eff} \ge 0,~ \rho^{eff}+p_t^{eff} +\frac{q^2}{4\pi\,r^4}\ge 0,~~~~~ \\
SEC &:& \rho^{eff}+p_r^{eff} \ge 0,~~\rho^{eff}+p_r^{eff}+2\,p_t^{eff} +\frac{q^2}{4\pi\,r^4}\ge 0,~~~~~\\
DEC &:& \rho^{eff}+\frac{q^2}{8\pi\,r^4}\ge 0, ~ \rho^{eff}-p_r^{eff}+\frac{q^2}{4\pi\,r^4}\ge 0,~\rho^{eff}-p_t^{eff}\ge 0,~~\\
TEC &:& \rho^{eff}-p_r^{eff}-2p_t^{eff}\ge 0.
\end{eqnarray}
These inequalities are known as energy conditions, where NEC means null energy condition, WEC weak energy condition, SEC strong energy condition, DEC dominant energy condition and TEC trace energy condition. It is worth mentioning that WEC and SEC include NEC and DEC contains WEC. From Fig. \ref{fig9} it is observed that all inequalities are satisfied for different values of $\beta$ and $\chi$. Moreover, the fact that the energy conditions are satisfied prevents non-physical properties, such as energy propagating faster than the speed of light, negative matter density or empty space spontaneously decaying into compensating regions of positive and negative energy.
%%%%%%%%%%%%%%%%%%%%

%%%%%%%%%%%%%%%%%%%%%%%%%%%%%%%%%%%%

\section{Concluding remarks}\label{sec10}
In the present paper we have investigated the nature of charged anisotropic
fluid spheres within the framework of $f(R,\mathcal{T})=R$ gravity theory. This work can be divided in three principal ingredients
\begin{itemize}
    \item Due to the fact that any perfect fluid solution to Einstein gravity theory satisfies also the same condition in $f(R,\mathcal{T})=R$ gravity (Eq. (\ref{diff})), we have extended the isotropic Durgapal-Fuloria model within modified gravity theories. Moreover, this fact is completely independent of the choice made about the $f(R,\mathcal{T})=R$ function and the Lagrangian density $\mathcal{L}_{m}$. In this opportunity we have taken $f(R,\mathcal{T})=R+2\chi\mathcal{T}$,  where $R$ and $\mathcal{T}$ are the Ricci scalar and the trace of energy-momentum tensor respectively, and $\chi$ is a dimensionless parameter. Furthermore, the versatility of the previous election on $f (R, \mathcal{T}$) has been tested several times in the study of compact structures.
    \item The second main point of this paper, concerns the inclusion of electric properties on the $f(R,\mathcal{T}$) Durgapal-Fuloria solution. This ingredient is characterized by two constant parameters $\chi$ and $\alpha$.
    \item Finally, the third principal feature is the extension of gravitational decoupling by minimal geometric deformation (MGD) approach in the arena of f (R,$\mathcal{T}$) gravity theory. This extension was plausible due two facts i) Bianchi's identities are satisfied and ii) is possible to do gravitational decoupling. As has been studied extensively in the literature, this method allows the extension of isotropic fluid solution to anisotropic domains. The anisotropic sector comes from adding an extra source $\theta_{\mu\nu}$ through a dimensionless coupling constant $\beta$. Of course, the presence of anisotropic matter distributions describe a more realistic scenario from the astrophysical point of view and also introduce many interesting and intriguing features that enhance equilibrium and stability mechanics in the stellar interior.
\end{itemize}
The charged anisotropic Durgapal-Fuloria model developed in this work, describes very well the physical and mathematical behaviour of compact structures like neutron or quark stars. As shown the main salient properties that characterize the model satisfy all the general requirements to ensure a well behaved system. These general requirements refer to   \begin{itemize}
    \item All the effective thermodynamic observables, \i.e matter density $\rho^{eff}$, radial pressure $p^{eff}_{r}$ and tangential pressure $p^{eff}_{t}$ are monotonic decreasing function with increasing radius, positive defined and finite, free from  mathematical singularities and with their maximum values attained at the center of the star.
    \item Vanished anisotropy factor $\Delta$ at the center, positive defined and increasing function towards the surface (in the case $p^{eff}_{t}>p^{eff}_{r}$). Respect to the electric intensity $E^{2}$, it must be null at the center and increasing function of the radial coordinate.
    \item Positive defined and well behaved energy-momentum tensor. It is checked studying energy conditions inside the star. \item Preservation of causality, \i.e both the radial and tangential subliminal sound speeds within the star must be lees than the speed of light (causality condition also can be checked by means of weak energy condition).
    \item Stable system under the presence of extra ingredients such as radial perturbations in the matter content due to the presence of anisotropies. This fact is investigated using for example Abreu's criterion (as in our case). This determines whether the equilibrium is stable or not. Furthermore, equilibrium analysis is performed applying TOV equation. TOV equation allows the study under different forces that act on the star, in this case the star is under four different forces, the hydrostatic $F_{h}$,  electric $F_{e}$, anisotropic $F_{a}$ and gravitational $F_{g}$. The first three forces counteract the gravitational one, allowing a more stable, compact and massive system.
\end{itemize}
All the criteria described above can be observed in Figs. ~\ref{fig1},~\ref{fig2}, ~\ref{fig3}, ~\ref{fig6}, ~\ref{fig7}, ~\ref{fig8} and ~\ref{fig9}. On the other hand, to obtain the complete set of physical quantities and physical parameters of the system, one needs to ensure a smooth joining between the interior geometry and the exterior space-time. It is worth mentioning that beyond the star all the matter content is null, then the appropriate exterior space-time is the Reissner-Nordstr\"{o}m one. The complete data obtained is shown in tables \ref{table1}, \ref{table2}, \ref{table3} and \ref{table4}. Table I exhibits the predicted radius, the surface redshift and the mass-radius ratio for different strange star candidates. The range $6.7<R<10.8$ of the predicted radius for these stars, is in complete agreement with many investigations on compact objects reported in the literature. The redshift values are the corresponding ones for compact object containing anisotropic matter distribution. As was pointed out by Ivanov ~\cite{r55}, in presence of anisotropies the maximum surface redshift is $Z_s<5.211$, it is clear that our model satisfies the above constraint. On the other hand, in the case of anisotropic charged spheres the mass-radius ratio is more general than its isotropic counterpart (Buchdahl's limit ~\cite{r83}). In this direction Andreasson ~\cite{r84} established the lower limit  and the upper bound was given by Bohmer and Harko ~\cite{r85}. So, the mass-radius ratio (compactness factor) values obtained are within the range established in the mentioned works. In tables \ref{table2}, \ref{table3} and \ref{table4}, it can be appreciated the effect of the different parameters introduced by  $f (R,\mathcal{T})$ theory, mathematical solution procedure to obtain the $e^{\nu(r)}$ metric potential and gravitational decoupling \i.e $\chi$, $\alpha$ and $\beta$ respectively, on the the star LMC X-4 ($M_{\bigodot}$=1.29). These effects can be summarized as follows
\begin{itemize}
    \item Table \ref{table2}: Fixing $\alpha$ and $\beta$ to $1.177$ and $0.1$ respectively and varying $\chi$ from $-0.2$ to $0.4$, the radius increases progressively, the central pressure, central density and surface density decrease progressively. In conclusion, taking smaller values of $\chi$ the object becomes more compact, dense and massive.
  
    \item Table \ref{table3}: Fixing $\alpha$ and $\chi$ to $1.177$ and $0.2$ respectively and varying $\beta$ from $0.05$ to $0.20$, we can see that all the physical parameters such as the central pressure, central density and surface density when $\beta$ increases. In distinction with what happens with the radius, which increases progressively when $\beta$ increases. It means, that the $\beta$ parameter introduced by gravitational decoupling method to includes the anisotropic behaviour of the matter distribution has a great influence on many physical parameters of the star.
    
    \item Table \ref{table4}: Fixing $\chi$ and $\beta$ to $0.2$ and $0.1$ respectively and varying $\alpha$ from $1.000$ to $1.400$, It is observed that the physical parameters have the same behavior as shown in table \ref{table2}.
\end{itemize}
Let us now discuss the effect of parameters $\chi$, $\alpha$ and $\beta$ on the compactness and surface redshift:
As we can see from Tables \ref{tab5} and \ref{tab7}, the surface redshift and compactness remain the same for fixed values of $\beta$. Since metric potential $\xi$ depends on $\beta$ not on $\xi$ and $\alpha$, then the compactness and surface redshift is not altered. On the other hand, From Table \ref{tab6} we observe that the surface redshift and compactness factor decrease when $\beta$ increases. Furthermore, tables \ref{tab5}, \ref{tab6} and \ref{tab7} contain the lower and upper values of the mass-radius ratio, as we can see the upper bound beats maximum Buchdahl's constraint. In brief, it is noted that taking small values for $\alpha$, $\beta$ and $\chi$ one obtains more compact  objects. \\
At this point we can highlight the similarities and differences between GR and  $f(R,\mathcal{T})=R$ theories. As was pointed out early in both theories the isotropic condition is the same, it means that under certain conditions (it will be clear early) any solution (uncharged isotropic or charged isotropic) that satisfies $f(R,\mathcal{T})=R$ field equations also satisfies Einstein's field equations. Although $f(R,\mathcal{T})=R$ theory was developed in the cosmological context, research on the existence of compact objects within this theory shows great success such as in GR. The inclusion of modifications in the geometric and matter sector opens the door to study more complex structures that can give answers to the limitations presented by GR. However, despite being an alternative to GR, in the limit $\chi\rightarrow0$ all the results predicted by GR must be recovered. In this case, if we set $\chi=0$ it is clear that Eq. (\ref{frt}) becomes $f(R,\mathcal{T})=R$, so from Eq. (\ref{1.5}) charged anisotropic Einstein-Maxwell field equations are obtained. On the other hand, if we set $\chi=0$ and $\beta=0$ charged isotropic solution is acquired. Furthermore, setting $\chi=0$ and $\alpha=1$ anisotropic Durgapal-Fuloria model is obtained in the arena of GR. Finally, if we set $\chi=0$, $\beta=0$ and $\alpha=1$ original Durgapal-Fuloria model is recovered.

\begin{acknowledgments}
S. K. Maurya acknowledge continuous support
and encouragement from the administration of University of Nizwa. F. Tello-Ortiz thanks the financial support of the project ANT-1855 at the Universidad de Antofagasta, Chile.
\end{acknowledgments}

%%%%%%%%%%%%%%%%%%%%%%%%%%%%%%%%%%%%%%%%%%%%%%%%%%%%
\end{document}